\documentclass[%
 aip,
 nofootinbib,
 amsmath,amssymb,
preprint,%
]{revtex4-1}

\usepackage{graphicx}% Include figure files
\usepackage{dcolumn}% Align table columns on decimal point
\usepackage{bm}% bold math
\usepackage[utf8]{inputenc}
\usepackage[T1]{fontenc}
\usepackage{mathptmx}
\usepackage{etoolbox}
\usepackage{algorithm, algpseudocode}

\makeatletter
\def\@email#1#2{%
 \endgroup
 \patchcmd{\titleblock@produce}
  {\frontmatter@RRAPformat}
  {\frontmatter@RRAPformat{\produce@RRAP{*#1\href{mailto:#2}{#2}}}\frontmatter@RRAPformat}
  {}{}
}%
\makeatother

\graphicspath{{Figures_RFF_kgEDMD/}}

\newcommand{\R}{\mathbb{R}}                                     % real numbers
\newcommand{\pd}[2]{\frac{\partial#1}{\partial#2}}              % partial derivatives
\newcommand{\pdtwo}[3]{\frac{\partial^2 #1}{\partial #2 \partial #3}}			% second order partial derivatives

\newcommand{\innerprod}[2]{\left\langle #1,\, #2 \right\rangle} % scalar product
\newcommand{\ts}{\hspace*{0.1em}}                               % thin space
\newcommand{\diff}{\mathrm{d}}                                   %'d' for integrals
\newcommand{\htop}{\mathrm{H}}                                  % conjugate transpose
\newcommand{\rff}{\mathrm{RFF}}                                 %RFF symbol
\newcommand\xqed[1]{\leavevmode\unskip\penalty9999 \hbox{}\nobreak\hfill \quad\hbox{#1}}
\newcommand{\exampleSymbol}{\xqed{$\triangle$}}

\DeclareMathOperator{\spn}{span}

\newcommand{\bX}{\mathbb{X}}
\newcommand{\bH}{\mathbb{H}}
\newcommand{\bE}{\mathbb{E}}

\newcommand{\cK}{\mathcal{K}}
\newcommand{\cL}{\mathcal{L}}
\newcommand{\cG}{\mathcal{G}}
\newcommand{\cA}{\mathcal{A}}
\newcommand{\cLO}{\mathcal{L}_{\mathrm{OD}}}

\newcommand{\A}{\mathbf{A}}

\newcommand{\K}{\mathbf{K}}
\newcommand{\M}{\mathbf{M}}

\newcommand{\U}{\mathbf{U}}
\newcommand{\W}{\mathbf{W}}
\newcommand{\G}{\mathbf{G}}
\newcommand{\X}{\mathbf{X}}

\newcommand{\T}{\mathbf{T}}
\newcommand{\LT}{\mathbf{L}}

\newtheorem{theorem}{Theorem}[section]

\newtheorem{example}[theorem]{Example}

\begin{document}

\title{Efficient Approximation of Molecular Kinetics using Random Fourier Features}
% Force line breaks with \\
\author{Feliks Nüske}
\email{nueske@mpi-magdeburg.mpg.de, S.Klus@hw.ac.uk}
 \affiliation{ 
Max-Planck-Institute for Dynamics of Complex Technical Systems, Magdeburg, Germany
}%
\author{Stefan Klus}%
\affiliation{ 
Heriot--Watt University, Edinburgh, United Kingdom
}%

\date{\today}% It is always \today, today,

\begin{abstract}
Slow kinetic processes of molecular systems can be analyzed by computing dominant eigenpairs of the Koopman operator or its generator. In this context, the Variational Approach to Markov Processes (VAMP) provides a rigorous way of discerning the quality of different approximate models. Kernel methods have been shown to provide accurate and robust estimates for slow kinetic processes, but are sensitive to hyper-parameter selection, and require the solution of large-scale generalized eigenvalue problems, which can easily become computationally demanding for large data sizes. In this contribution, we employ a stochastic approximation of the kernel based on random Fourier features (RFFs), to derive a small-scale dual eigenvalue problem which can easily be solved. We provide an interpretation of this procedure in terms of a finite randomly generated basis set. By combining the RFF approach and model selection by means of the VAMP score, we show that kernel parameters can be efficiently tuned, and accurate estimates of slow molecular kinetics can be obtained for several benchmarking systems, such as deca alanine and the NTL9 protein.
\end{abstract}

\maketitle

\section{Introduction}

The automated extraction of essential information about thermodynamic and kinetic properties of molecular systems from large-scale computer simulations remains one of the major open problems in computational physics and chemistry to this day. The challenge consists in processing very long time series of high-dimensional data in such a way that \emph{relevant} features of the molecular system are determined automatically, with limited user input or intervention. The precise definition of what is regarded as relevant depends on the context, but in many applications, the determination of \emph{slow} relaxation processes and their associated metastable sets is among the most important features to be detected. This applies, for example, to protein folding, ligand binding, or protein--protein association~\cite{Freddolino2010, Plattner2017, paul2017}.

Let us highlight two milestones that have shaped the field in the last two decades: The first was the inception of Markov State Models (MSMs) in the late 1990s and early 2000s~\cite{ulam1960,DELLNITZ1999, SCHUETTE1999}, which allowed for the construction of easily interpretable discrete kinetic models, from which a wealth of information could be extracted. The second milestone was the variational approach to conformational dynamics (VAC) for reversible systems~\cite{NOE2013, NUESKE2014}, and its extension called \emph{variational approach to Markov processes} (VAMP)~\cite{WU2020}, which provided a rigorously defined scalar objective function (called VAMP score) to assess the quality of kinetic models. Combined, these developments have led to a standard workflow for the construction of MSMs, usually consisting of a linear dimensionality reduction using \emph{time-lagged independent component analysis} (TICA)~\cite{PEREZ2013}, geometric clustering, as well as MSM estimation and validation~\cite{Prinz2011c,BowmanPandeNoe_MSMBook}. It was also shown that MSMs and the VAC are closely related to the Koopman operator framework~\cite{Koopman1931, Mezic2005}, and in particular the \emph{extended dynamic mode decomposition} (EDMD) algorithm~\cite{WILLIAMS2015, KLUS2016numerical}, which have primarily emerged in the analysis of fluid dynamics and related fields. The Koopman framework is therefore a unifying theme for the construction of kinetic models. Extensions of the EDMD algorithm can also be used to approximate the generator of dynamical systems (e.g., the Kolmogorov backward operator for stochastic differential equations) as shown in~\cite{Klus2020}.

Despite the remarkable success of MSMs, the construction process remains somewhat unsatisfactory, at least from a theoretical perspective. The initial TICA step presupposes that there is a kinetically informative linear subspace of the initial set of descriptors used for MSM building. In order to move beyond this requirement, algorithms based on deep neural networks (DNNs) were introduced in recent years~\cite{MARDT2018,Chen2019}, allowing to learn in one step the complete non-linear transformation from raw atomic coordinates to low-dimensional features that optimize the VAMP score. Both the MSM construction process (including state definition) and neural networks require a non-linear optimization step, which is typically solved by means of stochastic algorithms. A significant number of hyper-parameters need to be tuned in both cases, such as network architecture, network depth and height for DNNs, or number of cluster centers, number of iterations, convergence threshold, etc.\ for clustering-based MSMs.

A different avenue that has been considered is to employ reproducing kernels, which have long been known in machine learning~\cite{WENDLAND04, Steinwart2008:SVM} as a powerful non-linear model class with rich theoretical properties. Kernel methods have also been demonstrated in many applications to be fairly robust if the available data is only small- to medium-sized. Kernel-based versions of EDMD were suggested in~\cite{WILLIAMS2015b, Klus:2020aa, Klus2020a}, and applied to high-dimensional molecular simulations in~\cite{KBSS18}. Extracting slow timescales from kernel representations requires the solution of a large-scale generalized eigenvalue problem, where the size of the matrix depends on the number of snapshots. Even for moderate data size, the solution of this problem can be challenging. Moreover, although kernel methods usually require only a few hyper-parameters (such as the bandwidth of the hugely popular Gaussian kernel), these need to be tuned accurately, requiring to solve the above-mentioned eigenvalue problem many times, which can be prohibitively time-consuming.

In this work, we show that kernel methods can be made tractable for the analysis of molecular kinetics by employing \emph{Random Fourier Features} (RFFs)~\cite{rahimi2007RFF}, a well-known low-rank approximation technique for kernel matrices, to convert the original large-scale eigenvalue problem into a much smaller one, which can be easily solved. We derive the associated small-scale matrix problem both for the Koopman operator and its generator, and show that this approach can be interpreted as applying EDMD with a specific randomly chosen basis set. We demonstrate that the VAMP score can be effectively used for hyper-parameter tuning of the kernel method. The combination of these ideas leads to an efficient approximation algorithm, which is robust with respect to the data size and the stochastic Fourier feature selection. Therefore, our method presents a promising new direction for the automated analysis of molecular kinetics.

The remainder of this paper is structured as follows: In Section~\ref{sec:basics}, we review elements of the Koopman formalism, the EDMD approximation algorithm, reproducing kernel Hilbert spaces (RKHSs), and Koopman modeling using RKHSs. Then, in Section~\ref{sec:main_contribution}, we present our main contribution, an approximation algorithm for the Koopman operator or generator based on Random Fourier Features. Further computational details, especially regarding model selection, are discussed in Section~\ref{sec:methods}, while applications to benchmarking problems in molecular dynamics are shown in Section~\ref{sec:results}.

\section{Theoretical Basics}
\label{sec:basics}

We will briefly introduce the stochastic Koopman operator and its generator as well as numerical methods to estimate these operators and their spectral properties from data.

\subsection{Stochastic Dynamical Systems}

Computer simulations are routinely used to explore the thermodynamics and kinetics of molecular systems. Many available implementations of molecular dynamics (MD) can be described as a stochastic process $\{ X_t \}_{t \ge 0}$, where $t $ is the time and $X_t \in \bX \subset \R^d$ the state of the system at time $t$. In particular, a widely used model in this context is given by \emph{stochastic differential equations} (SDEs) of the form
\begin{equation}
    \label{eq:SDE}
    \diff X_t = F(X_t)\,\diff t + G(X_t)\,\diff W_t.
\end{equation}
where $F \colon \R^d \to \R^d$ and $G\colon \R^d \to \R^{d\times d}$ are vector- and matrix-valued fields, respectively, called the \emph{drift} and \emph{diffusion} of the SDE, and $W_t$ is $d$-dimensional Brownian motion. The covariance matrix of the diffusion is denoted by
\begin{equation*}
    a(x) := G(x) \ts G(x)^\top.
\end{equation*}
As far as the thermodynamics is concerned, the primary goal of molecular simulations is to sample the \emph{Boltzmann distribution}
\[ \diff \mu(x) \sim \exp(-\beta V(x)) \, \diff x, \]
where $V$ is the potential energy at position $x$ and $\beta^{-1} = k_B T$ is the inverse temperature. Most simulation protocols are built to achieve this goal, i.e., the Boltzmann distribution is invariant, and long time averages converge to spatial averages with respect to $\mu$ (ergodicity). A standard example is \emph{Brownian} or \emph{overdamped Langevin dynamics}
\begin{equation}
\label{eq:od_langevin}
    \diff X_t = -\frac{1}{\gamma}\nabla V(X_t) \, \diff t  + \sqrt{2\gamma^{-1} \beta^{-1}} \, \diff W_t,
\end{equation}
with friction parameter $\gamma$.

\subsection{Koopman Operator and Generator}

When it comes to inferring kinetic information about $X_t$, significant progress has been made in recent years by considering an operator-based approach to the statistics of the stochastic process $X_t$. For a fixed lag time $t \geq 0$, the \emph{Koopman operator}~\cite{Koopman1931} $\cK^t$ acts on a functions $\phi \colon \mathbb{X} \to \R$ by taking the conditional expectation:
\begin{equation*}
    %\label{eq:def_koopman}
    \cK^t \phi(x) = \bE^x[\phi(X_t)] = \bE[\phi(X_t) \mid X_0 = x].
\end{equation*}
In other words, evaluation of the function $\cK^t \phi$ at $x$ yields the expectation of $\phi$ after starting the dynamics at $x$, and evolving until time $t$. The linear operators $\cK^t$ satisfy the semigroup equation $\cK^{s+t} = \cK^{s}\cK^{t}$, which implies the existence of another linear operator $\cL$, called \emph{Koopman generator}, such that the following linear differential equation holds for the function $\phi_t = \cK^t \phi$:
\begin{equation*}
    %\label{eq:heat_eq_generator}
    \pd{}{t} \cK^t \phi(x) = \cL \cK^t \phi(x).
\end{equation*}
For SDE dynamics of the form~\eqref{eq:SDE},  stochastic calculus shows that $\cL$ is a second-order differential operator, and we obtain the \emph{Kolmogorov equation}
\begin{equation*}
    %\label{eq:backward_kolm}
    \begin{split}
    \pd{}{t}\phi_t(x) &= \sum_{i=1}^d F(i)(x) \pd{}{x^i}\phi_t(x) + \frac{1}{2} \sum_{i, j=1}^d a(i,j)(x) \pdtwo{}{x^i}{x^j}\phi_t(x) \\
    &= F(x) \cdot \nabla \phi_t(x) + \frac{1}{2} a(x) : \nabla^2 \phi_t(x) \\ &= \cL \phi_t(x),
    \end{split}
\end{equation*}
where we used the colon notation for the Frobenius inner product between matrices, i.e., $A : B = \sum_{i,j} A(i, j) B(i, j)$. For overdamped Langevin dynamics, the generator is given by
\begin{align*}
    \cLO &= -\frac{1}{\gamma}\nabla V \cdot \nabla + \gamma^{-1}\beta^{-1} \Delta.
\end{align*}

We will summarily refer to the Koopman operators or the generator as \emph{dynamical operators} for the dynamical system $X_t$. If the dynamics are reversible with respect to the invariant distribution $\mu$, then the generator $\cL$ is symmetric with respect to the inner product
\begin{equation*}
    \innerprod{\phi_1}{\phi_2}_\mu = \int_\bX \phi_1(x) \ts \phi_2(x)\,\diff \mu(x).
\end{equation*}
As a consequence, all eigenvalues of the generator $\cL$ are real-valued and non-positive, i.e., for all solutions of the equation
\begin{equation*}
    %\label{eq:ev_eq_generator}
    \cL \psi_i = \kappa_i \ts \psi_i,
\end{equation*}
we have $\kappa_i \leq 0$. An object of prime interest for molecular kinetics are low-lying eigenvalues $0 \leq -\kappa_i \ll 1$, as these encode slow motions corresponding to metastable behavior encountered in many examples of molecular systems~\cite{Davies1982,SCHUETTE1999}. The connection between close-to-zero eigenvalues and metastability is also apparent in the fact that eigenfunctions of the generator are also eigenfunctions of the Koopman operator at any $t$, with eigenvalue $\lambda_i(t) = e^{-\kappa_i t}$. If $-\kappa_i \ll 1$, then the corresponding eigenvalues $\lambda_i(t)$ decay slowly, as expected for metastable dynamics.

\subsection{Finite-dimensional Approximation}
\label{subsubsec:finite_dim_approx}
The basic data-driven approximation algorithm for the Koopman operator is called \emph{extended dynamic mode decomposition} (EDMD)~\cite{WKR15}, its adaptation to the generator is then called \emph{gEDMD}~\cite{Klus2020}. Both methods require choosing a finite \emph{dictionary} of observable functions $\{\phi_1, \ldots, \phi_n\}$. The projections of the dynamical operators onto the linear span of these functions can be represented by matrices:
\begin{align*}
    \K^t &= \G^{-1}\A^t, & \LT &= \G^{-1}\A^L,
\end{align*}
with mass and stiffness matrices given by
\begin{align*}
     \G(i, j) &= \innerprod{\phi_i}{\phi_j}_\mu, & \A^t(i, j) &= \innerprod{\phi_i}{\cK^t \phi_j}_\mu, & \A^L(i, j) &= \innerprod{\phi_i}{\cL \phi_j}_\mu.
\end{align*}
Given a long trajectory $\{x_1, x_2, \dots, x_{m+1}\}$ of the process $X_t$, the data-based estimators for these matrices are given by
\begin{align*}
%\label{eq:gEDMD}
    \hat{\G}(i, j) &= \frac{1}{m}\sum_{l=1}^m \phi_i(x_l) \ts \phi_j(x_l), & \hat{\A}^t(i, j) &= \frac{1}{m}\sum_{l=1}^m \phi_i(x_l) \ts \phi_j(x_{l+1}), \\ \nonumber \hat{\A}^L(i, j) &= \frac{1}{m}\sum_{l=1}^m \phi_i(x_l) \ts \cL\phi_j(x_l),
\end{align*}
where we assumed the time step between samples to equal $t$ in the formula for $\hat{\A}^t$. This is not required in practice if a sliding-window estimator is used, see~\cite{BowmanPandeNoe_MSMBook,Klus:2018ab} for details.

For reversible dynamics, the stiffness matrix $\A^L$ and its estimator $\hat{\A}^L$ above can be replaced by the following energy-like expressions~\cite{Klus2020}:
\begin{equation}
\begin{split}
    \A^L(i, j) &= -\frac{1}{2}\int_\bX \nabla^\top \phi_i(x) \ts a(x) \nabla \phi_j(x) \, \diff\mu(x), \\
    \hat{\A}^L(i, j) &= -\frac{1}{2m} \sum_{l=1}^m \nabla^\top \phi_i(x_l) \ts a(x_l) \nabla \phi_j(x_l).
\end{split} \label{eq:A_rev}
\end{equation}
The latter estimator only requires first-order derivatives and retains the symmetry of $\A^L$ even for finite data.

Eigenvalues of the dynamical operators can be approximated by diagonalizing the data-driven estimators $\hat{\K}^t = \hat{\G}^{-1}\hat{\A}^t$ and $\hat{\LT} = \hat{\G}^{-1}\hat{\A}^L$, or equivalently, by solving one of the generalized eigenvalue problems
\begin{align}
\label{eq:edmd_ev_problem}
    \hat{\A}^t \mathbf{v}_i &= \hat{\lambda}_i(t) \ts \hat{\G} \ts \mathbf{v}_i, & \hat{\A}^L \mathbf{v}_i &= \hat{\kappa}_i \ts \hat{\G} \ts \mathbf{v}_i.
\end{align}

\subsection{Collective Variables}
\label{subsubsec:cg_generator}

For large molecular systems, the dictionary is typically defined in terms of a set of lower-dimensional descriptors $z \in \R^N$, $N \leq d$, often called \emph{collective variables} (CVs), such as intramolecular distances or angles. Formally, CVs are given by a smooth mapping $\xi \colon \bX \to \R^N$. Choosing a dictionary $\{\tilde{\phi}_i\}_{i=1}^n$ of functions depending only on CV space, that is $\tilde{\phi}_i(x) = \phi_i(\xi(x))$, does not lead to any conceptual changes to the EDMD approach. When approximating the generator, the gEDMD method is straightforwardly adapted by computing all $x$-derivatives in the definition of the generator using the chain rule. This leads to a particularly elegant reformulation of the symmetric estimator~\eqref{eq:A_rev} in terms of an effective diffusion matrix
\begin{equation}
\label{eq:effective_diffusion}
    a^\xi(x) = \nabla_x^\top \xi(x) \ts a(x) \nabla_x \ts \xi(x),
\end{equation}
where $\nabla_x \ts \xi$ is the $d\times N$-dimensional Jacobian of the mapping $\xi$. The symmetric estimator~\eqref{eq:A_rev} then becomes
\begin{equation*}
    \hat{\A}^L(i, j) = -\frac{1}{2m} \sum_{l=1}^m \nabla_z^\top \phi_i(z_l) a^\xi(x_l) \nabla_z \phi_j(z_l)
\end{equation*}
as shown in~\cite{Klus2020}. Note that once the effective diffusion matrices have been computed, only the basis functions and their derivatives in CV-space are required.

\subsection{Reproducing Kernel Hilbert Space}

A critical modeling decision for the (g)EDMD approach is the choice of the dictionary $\{\phi_1, \ldots, \phi_n\}$. Recent works~\cite{WILLIAMS2015b,Klus:2020aa,KBSS18,Klus2020a} have shown that accurate results, even for large systems, can be obtained using methods based on \emph{reproducing kernel Hilbert spaces} (RKHS) generated by a kernel function $k$. These approaches lead to powerful approximation spaces, yet only depend on few model parameters.

A reproducing kernel Hilbert space (RKHS)~\cite{Aronszajn50,Steinwart2008:SVM} is a Hilbert space $\bH$, with inner product $\innerprod{\cdot}{\cdot}_\bH$, of continuous functions on $\bX$, defined by a symmetric, positive-definite two-argument function $k(x, y)$, called the \emph{kernel}. Each such kernel corresponds to a unique RKHS~\cite{Aronszajn50}. Given $k$, one can define a map from $\bX$ into $\bH$, called the \emph{feature map}, by
\[ \Phi(x)(y) = k(x, y). \]
Functions $f$ in the RKHS then satisfy the reproducing property
\begin{equation*}
    \innerprod{f}{\Phi(x)}_\bH = f(x) \quad \forall f \in \bH.
\end{equation*}

\begin{example}
Arguably the most widely used kernel is the Gaussian radial basis function (RBF) kernel
\begin{equation*}
    %\label{eq:gaussian_kernel}
    k_\sigma(x, y) = \exp\left(-\frac{1}{2\sigma^2}\|x - y\|^2\right),
\end{equation*}
with bandwidth parameter $\sigma > 0$. On a periodic domain $[-L, L]^d$, the closely related periodic Gaussian kernel
\begin{equation*}
    %\label{eq:periodic_gaussian}
    k^p_\sigma(x, y) = \exp\left(-\frac{2}{\sigma^2}\sum_{i=1}^d \sin^2(0.5(x^i - y^i))\right),
\end{equation*}
analogously generates an RKHS of periodic functions~\cite{mackay1998GP}. \exampleSymbol
\end{example}
For many popular kernels, including the Gaussian RBF kernel, the associated RKHS is infinite-dimensional, and densely embedded into most relevant function spaces. Kernels therefore provide powerful approximation spaces, reducing the problem of dictionary selection to the choice of a single function, the kernel, and its hyper-parameters, such as the bandwidth for the Gaussian RBF kernel.

\subsection{RKHS Representation of Dynamical Operators}
\label{subsec:rkhs_dyn_op}

Given a kernel $k$ with RKHS $\bH$, the analogues of the matrices $\G$, $\A^t$ and $\A^L$ in Section~\ref{subsubsec:finite_dim_approx} are given by linear operators on $\bH$, see~\cite{Klus:2020aa,Klus2020a} for a derivation:
\begin{align}
\label{eq:def_rkhs_ops}
\mathcal{G} f &= \int_\mathbb{X} f(x) \Phi(x)\,\mathrm{d}\mu(x),
& \mathcal{A}^t f &= \int_\mathbb{X} \cK^t f(x) \Phi(x)\,\mathrm{d}\mu(x),
& \mathcal{A}^L f &= \int_\mathbb{X} \cL f(x) \Phi(x)\,\mathrm{d}\mu(x).
\end{align}
Here, $\Phi$ is again the feature map. The eigenvalue problem~\eqref{eq:edmd_ev_problem} then turns into an infinite-dimensional generalized eigenvalue equation for eigenfunctions $\psi_i \in \bH$:
\begin{align}
\label{eq:eigenvalue_prob_rkhs}
    \cA^t \psi_i &= \lambda_i(t) \cG \psi_i, & \cA^L \psi_i &= \kappa_i \cG \psi_i.
\end{align}
Natural empirical estimators for the RKHS operators in~\eqref{eq:def_rkhs_ops} are given by
\begin{align*}
%\label{eq:emp_rkhs_ops}
    \hat{\mathcal{G}} f &= \frac{1}{m} \sum_{l=1}^m f(x_l) \Phi(x_l), &
    \hat{\mathcal{A}}^t f &= \frac{1}{m} \sum_{l=1}^m f(x_{l+1}) \Phi(x_l), &
    \hat{\mathcal{A}}^L f &= \frac{1}{m} \sum_{l=1}^m \mathcal{L}f(x_l) \Phi(x_l).
\end{align*}
The range of these operators is in the linear span of the feature map at the data sites, $\bH_m = \spn\{\Phi(x_l)\}_{l=1}^m$. Therefore, it makes sense to restrict these operators to the $m$-dimensional space $\bH_m$, which leads to the matrix representations
\begin{equation}
\begin{split}
\label{eq:kernel_matrices}
    \left[\K_{X}\right](r, s) &= k(x_r, x_s), \\[1.5ex]
    \left[\K_{X}^t\right](r, s) &= k(x_{r+1}, x_s),\\
    \left[\K^L_{X}\right](r, s) &= \frac{1}{2} \sum_{i,j=1}^d a(i, j)(x_r) \pdtwo{}{x^i}{x^j} k(x_r, x_s) + \sum_{i=1}^d F_i(x_r) \pd{}{x^i} k(x_r, x_s),
\end{split}
\end{equation}
and~\eqref{eq:eigenvalue_prob_rkhs} can be replaced by a matrix generalized eigenvalue problem
\begin{align}
\label{eq:ev_problem_kernel}
    \K^t_{X} \mathbf{w}_i &= \hat{\lambda}_i(t) \K_{X} \mathbf{w}_i, & \K^L_{X} \mathbf{w}_i &= \hat{\kappa}_i \K_{X} \mathbf{w}_i.
\end{align}
The derivatives in the definition of $\K_X^L$ are taken with respect to first argument of the kernel function. It is important to note that, as expected for a kernel method, assembling the matrices~\eqref{eq:kernel_matrices} requires only kernel evaluations at the data sites, along with kernel derivatives and coefficients of the SDE in the generator setting. For reversible dynamics, there is again an alternative symmetric formulation, which leads to the following Hermitian generalized eigenvalue problem instead of~\eqref{eq:ev_problem_kernel}:
\begin{equation} \label{eq:kernel_ev_rev}
\begin{split}
    \K^\mathrm{rev}_{X} \mathbf{w}_i &= \hat{\kappa}_i \K_{X} \K_{X} \mathbf{w}_i, \\
    \K^\mathrm{rev}_{X}(r,s) &= -\frac{1}{2m} \sum_{l=1}^m \nabla^\top k(x_l, x_r)a(x_l)\nabla k(x_l, x_s).
\end{split}
\end{equation}
The derivations of~\eqref{eq:kernel_matrices},~\eqref{eq:ev_problem_kernel} and~\eqref{eq:kernel_ev_rev} were shown in~\cite{Klus:2020aa,Klus2020a}, we repeat them for the reader's convenience in Appendix~\ref{app:derivation_kernel_ev}.

The linear problems~\eqref{eq:ev_problem_kernel} and~\eqref{eq:kernel_ev_rev} represent large-scale generalized eigenvalue problems. Solving these problems efficiently for large data sets is one of the central challenges common to most applications of kernel methods. In the following section, we present an approach based on a randomized low-rank representation obtained from Fourier transformation of the kernel function.

\section{Low-Rank Representations of Kernel Eigenvalue Problems}
\label{sec:main_contribution}

In this section, we will present our main contribution: an approach to solving the generalized eigenvalue problems~\eqref{eq:ev_problem_kernel} and~\eqref{eq:kernel_ev_rev} using a randomized low-rank decomposition.

\subsection{Random Fourier Features}
\label{subsec:rff}

\emph{Random Fourier Features} (RFFs) have been introduced in~\cite{rahimi2007RFF} as a low-rank approximation framework for large kernel matrices. The basis for this approach is Bochner's theorem, stating that a translation-invariant kernel $k$, which is normalized to $k(x, x) = 1$ for all $x$, possesses the following stochastic representation~\cite{bochner1933}:
\begin{equation*}
%\label{eq:rff_kernel}
    k(x, y) = \frac{1}{\sqrt{2\pi}}\int_{\mathbb{R}^d} e^{-i (x - y)^\top \omega}\,\mathrm{d}\rho(\omega) = \mathbb{E}^{\omega \sim\rho} \left[e^{-i x^\top \omega} \ts \overline{e^{-i y^\top \omega}} \right].
\end{equation*}
Here, $\rho$ is a probability measure in frequency space, called the spectral measure. This representation extends to the derivatives of the kernel function, namely
\begin{equation}
\label{eq:diff_rff_kernel}
    D^\alpha k(x, y) = \mathbb{E}^{\omega \sim\rho} \left[(-i\omega)^\alpha e^{-i x^\top \omega} \ts \overline{e^{-i y^\top \omega}} \right],
\end{equation}
where $\alpha \in \mathbb{N}^d$ is a multi-index and $\omega^\alpha = (\omega^1)^{\alpha(1)} \dots (\omega^d)^{\alpha(d)}$. The derivatives in the above expression are again taken with respect to the first argument. For the Gaussian RBF kernel with bandwidth $\sigma$, the spectral measure is also Gaussian with bandwidth $\sigma^{-1}$. For the periodic Gaussian kernel, the spectral measure is discrete, supported on all wave vectors $\frac{\pi}{L}\mathbf{k}$ for $\mathbf{k}\in \mathbb{Z}^d$, with probabilities
\[ \rho\left(\frac{\pi}{L}\mathbf{k}\right) = \prod_{j=1}^d I_{\mathbf{k}(j)}(\sigma^{-2}) e^{-\sigma^{-2}}, \]
where $I_k$ is the modified Bessel function of the first kind, see~\cite{tompkins2018}. Thus, drawing independent samples from the spectral measure is easily accomplished for many popular kernel functions.

\subsection{Representation of Kernel Matrices}

We apply the RFF representation to the generalized eigenvalue problems~\eqref{eq:ev_problem_kernel} and~\eqref{eq:kernel_ev_rev} by substituting it into the kernel matrices $\K_X, \K_X^t, \K_X^L$ given by~\eqref{eq:kernel_matrices}. We again denote samples in real space (in the form of a long trajectory\footnote{Instead of using one long trajectory, it would also be possible to extract a training data set $\{(x_l, y_l)\}_{l=1}^m$ from many short trajectories, see \cite{KNKWKSN18} for details.}) by $\{x_l\}_{l=1}^{m+1}$. Additionally, we assume there are $p$ samples $\{\omega_u\}_{u=1}^p$ in frequency space, sampled from the spectral measure $\rho$. Replacing expectation values by finite-sample averages, we obtain
\begin{equation}
\label{eq:rff_rep_kernel_mat}
\begin{split}
    \K_X &= \left[k(x_r, x_s) \right]_{r,s} = \left[\mathbb{E}^{\omega \sim \rho}\left[ e^{-i x_r^\top \omega} \ts \overline{e^{-i x_s^\top \omega}}\right] \right]_{r,s} \approx \frac{1}{p}\left[\M\M^\htop \right]_{r,s}, \\
    \K_X^t &= \left[k(x_{r+1}, x_s) \right]_{r,s} \approx \frac{1}{p}\left[\M^t \M^\htop \right]_{r,s},
\end{split}
\end{equation}
with RFF feature matrices
\begin{align}
\label{eq:rff_matrices}
\M &= \left[e^{-i x_r^\top \omega_u}\right]_{r,u}, & \M^t &= \left[e^{-i x_{r+1}^\top \omega_u}\right]_{r,u}.
\end{align}
That is, the rows of $ \M $ correspond to the data points and the columns of $ \M $ to the randomly sampled features. When approximating the generator, we use the representation of derivatives given in~\eqref{eq:diff_rff_kernel}, to find
\begin{align}
    \K_X^L &= \left[\frac{1}{2} \sum_{i,j=1}^d a(i, j)(x_r) \pdtwo{}{x^i}{x^j} k(x_r, x_s) + \sum_{i=1}^d F(i)(x_r) \pd{}{x^i} k(x_r, x_s)\right]_{r, s} \nonumber \\
    &\approx \frac{1}{p}\sum_{u=1}^p \left[\left(-\frac{1}{2} \sum_{i,j=1}^d a(i, j)(x_r) \omega_{u}(i)\omega_{u}(j) - i \sum_{i=1}^d F(i)(x_r) \omega_u(i) \right) e^{-i x_r^\top \omega_u} \ts \overline{e^{-i x_s^\top \omega_u}} \right]_{r, s} \nonumber \\
    &= \frac{1}{p} \left[ \M^L \M^\htop \right]_{r, s}, \label{eq:rff_matrix_generator}
\end{align}
with generator feature matrix $\M^L$, which can be written compactly using the Frobenius inner product as
\begin{equation*}
    \M^L = \left[\left(-\frac{1}{2}a(x_r) : (\omega_u \otimes \omega_u) - i F(x_r)\cdot \omega_u\right) e^{-i x_r^\top \omega_u}\right]_{r,u}.
\end{equation*}
We will address the reversible case further below after analyzing the RFF-based approximation in more detail. Together,~\eqref{eq:rff_rep_kernel_mat} and~\eqref{eq:rff_matrix_generator} provide low-rank approximations of the kernel matrices which can be easily assembled by evaluation of complex exponentials.

\subsection{Solution of the Generalized Eigenvalue Problems}
\label{subsec:solution_gev}

Let us now use the approximations~\eqref{eq:rff_rep_kernel_mat} and~\eqref{eq:rff_matrix_generator} to solve the generalized eigenvalue problems~\eqref{eq:ev_problem_kernel}, which read (note that the normalization $\frac{1}{p}$ can be omitted):
\begin{align}
\label{eq:ev_problem_rff}
    \M^t \M^\htop \mathbf{w}_i &= \hat{\lambda}_i(t) \M \M^\htop \mathbf{w}_i, &
    \M^L \M^\htop \mathbf{w}_i &= \hat{\kappa}_i \M \M^\htop \mathbf{w}_i.
\end{align}
If $p \leq m$, we can employ a standard trick to express this equivalently as a lower-dimensional eigenvalue problem. Defining $\mathbf{v}_i = \M^\htop \mathbf{w}_i$, we can verify directly that
\begin{align*}
    \M^\htop \M^t \mathbf{v}_i &= \M^\htop \M^t \M^\htop \mathbf{w}_i = \hat{\lambda}_i(t) \ts \M^\htop \M \M^\htop \mathbf{w}_i = \hat{\lambda}_i(t) \ts \M^\htop \M \mathbf{v}_i.
\end{align*}
The last equality shows that all non-zero eigenvalues of~\eqref{eq:ev_problem_rff} can be equivalently computed by the lower-dimensional \emph{dual} problem:
\begin{align}
\label{eq:ev_problem_transposed}
    \M^\htop \M^t \mathbf{v}_i &= \hat{\lambda}_i(t) \M^\htop \M \mathbf{v}_i, &
    \M^\htop \M^L \mathbf{v}_i &= \hat{\kappa}_i \M^\htop \M \mathbf{v}_i.
\end{align}
This $p$-dimensional generalized eigenvalue equation is the central problem to be solved within the context of RFF-based kernel approximation. If $p$ is much smaller than $m$,~\eqref{eq:ev_problem_transposed} already leads to computational savings compared to~\eqref{eq:ev_problem_kernel}.

In fact, it is not even necessary to assemble the matrices in~\eqref{eq:ev_problem_transposed}, because the equation can be solved by operating only on the RFF feature matrices $\M, \, \M^t, \,\M^L$. To this end, we apply a whitening transformation, which is widely used within the context of (g)EDMD. Using a truncated singular value decomposition of $\M$,
\begin{equation*}
    \M \approx \U \Sigma \W^\htop,
\end{equation*}
and retaining $r \leq p$ components of the SVD, we obtain a linear transformation $\T = \W \Sigma^{-1}$, which eliminates the matrix on the right-hand sides of~\eqref{eq:ev_problem_transposed}:
\begin{align*}
    \T^\htop \M^\htop \M \T = \Sigma^{-1} \W^\htop \W \Sigma \U^\htop \U \Sigma \W^\htop \W \Sigma^{-1} = \mathrm{Id}_{r \times r}.
\end{align*}
Applying the same transformation to the left-hand sides of~\eqref{eq:ev_problem_transposed} leads to the reduced matrices
\begin{align}
\label{eq:reduced_matrix}
    \mathbf{R}^t &= \T^\htop \M^\htop \M^t \T = \U^\htop \M^t \W \Sigma^{-1}, &
    \mathbf{R}^L &= \T^\htop \M^\htop \M^L \T = \U^\htop \M^L \W \Sigma^{-1}.
\end{align}

It is then sufficient to diagonalize these reduced matrices in order to solve~\eqref{eq:ev_problem_transposed}. If the truncation rank of $\M$ is significantly smaller than $p$, additional computational savings can be harnessed this way. Algorithm~\ref{alg: rff_spectral} summarizes our solution method in compact form.

\subsection{Interpretation as (g)EDMD Problem}

The transformation of the dual problem~\eqref{eq:ev_problem_transposed} to its reduced form is completely analogous to what is called the whitening transformation for (g)EDMD~\cite{Klus:2018ab}. This is not a coincidence: consider a finite dictionary defined by
\begin{equation*}
    \phi_{\rff}(x) := \begin{bmatrix} \phi_1(x) \\ \vdots \\ \phi_p(x)\end{bmatrix}
    = \begin{bmatrix} e^{i x^\top \omega_1} \\ \vdots \\ e^{i x^\top \omega_p} \end{bmatrix}.
\end{equation*}

Using the notation of Section~\ref{subsubsec:finite_dim_approx}, the empirical mass and stiffness matrices for this basis set are given by
\begin{align*}
\hat{\G}(u, v) &= \frac{1}{m}\sum_{l=1}^m \phi_u(x_l) \ts \overline{\phi_v(x_l)}
= \frac{1}{m}\sum_{l=1}^m e^{i x_l^\top \omega_u} e^{-i x_l^\top \omega_v} = \frac{1}{m}\left[\M^\htop\M\right](u, v),\\
\hat{\A}^t(u, v) &= \frac{1}{m}\left[\M^\htop\M^t \right](u, v), \\
\hat{\A}^L(u, v) &= \frac{1}{m} \left[\M^\htop\M^L \right](u, v),
\end{align*}
which are precisely the constituent matrices in~\eqref{eq:ev_problem_transposed}\footnote{We need to consider complex-valued observables at this point, leading to complex conjugation of the second argument in inner products.}. We conclude that the solution of the transposed problem is equivalent to the eigenvalue estimation based on (g)EDMD using a random dictionary of complex plane waves. With this interpretation at hand, we can also derive a convenient symmetric formulation for reversible dynamics, upon replacing the estimator for $\hat{\A}^L$ by its symmetric version~\eqref{eq:A_rev}:
\begin{equation*}
\begin{split}
    \hat{\A}^L(u, v) &= -\frac{1}{2m} \sum_{l=1}^m \nabla^\top \phi_u(x_l) a(x_l) \nabla \phi_v(x_l) \\
    &= -\frac{1}{2m} \sum_{l=1}^m \omega_u^\top a(x_l) \omega_v \, e^{i x_l^\top \omega_u} e^{-i x_l^\top \omega_v}.
\end{split}
\end{equation*}
To compute the reduced matrix, we just have to apply the transformation $\T$ from Section~\ref{subsec:solution_gev} from left and right, i.e.,
\begin{equation}
    \label{eq:reduced_matrix_rev}
    \mathbf{R}^L = \T^\htop \hat{\A}^L \T = \Sigma^{-1} \W^\htop \hat{\A}^L \W \Sigma^{-1},
\end{equation}
resulting again in a symmetric estimator.

\begin{algorithm}[H]
  \caption{RFF-based Spectral Approximation of the Operator or Generator}
  \label{alg: rff_spectral}
  \setlength{\tabcolsep}{.5ex}
  \begin{tabular}{ll}
    \textbf{Input:} & data matrix $\X = [x_1, \dots, x_{m+1}] \in \R^{d \times (m+1)}$, \\
    & kernel function $k$ with spectral measure $\rho$, \\
    & number of features $p$, truncation rule for singular values.\\
    \textbf{Output:} & Approximate eigenpairs $(\hat{\lambda}_i(t), \hat{\psi}_i)$ or $(\hat{\kappa}_i, \hat{\psi}_i)$ of the dynamical operator.
  \end{tabular}
  \hrule\vspace{0.2cm}
  \begin{algorithmic}[1]
    \State Draw $p$ samples $\{\omega_u\}_{u=1}^p$ from the spectral measure $\rho$.
    \State Form matrix $\M$ given by~\eqref{eq:rff_matrices}.
    \State Non-reversible case: form matrix $\M^t$ or $\M^L$ as in~\eqref{eq:rff_matrices} or~\eqref{eq:rff_matrix_generator}.
    \State Compute SVD of $\M$, choose rank $r$ according to truncation rule:
    $\M \approx \U \Sigma \W^\htop$.
    \State Form reduced matrix $\mathbf{R}^t$ or $\mathbf{R}^L$ according to~\eqref{eq:reduced_matrix} or~\eqref{eq:reduced_matrix_rev}.
    \State Compute eigenpairs of reduced problem $\mathbf{R}^t\mathbf{u}_i = \hat{\lambda}_i(t) \mathbf{u}_i$ or $\mathbf{R}^L\mathbf{u}_i = \hat{\kappa}_i \mathbf{u}_i$.
    \State Transform to original RFF basis: $\mathbf{v}_i = \mathbf{T} \mathbf{u}_i$, $\hat{\psi}_i(x) = \mathbf{v}_i^\htop \phi_\rff(x).$
  \end{algorithmic}
\end{algorithm}

\subsection{Computational Effort}

Let us break down the computational effort required for Algorithm~\ref{alg: rff_spectral}: The SVD of $\M$ requires $\mathcal{O}(mp^2)$ operations. Assembling the reduced matrix costs $\mathcal{O}(rpm)$ operations for the Koopman operator. In the case of generator approximation, we obtain an additional dependence on the dimension, due to the required contraction of derivatives of the kernel function. We can upper-bound the resulting effort as $\mathcal{O}(mp (d^2 + r))$ in the non-reversible case and by $\mathcal{O}(m(rpd + r^2d^2))$ in the reversible case. Note that this bound assumes a dense and state-dependent diffusion field $a$, it reduces to linear dependence on $d$ for diagonal diffusions, such as overdamped Langevin dynamics. Finally, the diagonalization of the reduced matrix amounts to $\mathcal{O}(r^3)$ floating point operations. Comparing this to full kernel method, we see that the most drastic computational savings will result from being able to choose $p$ small compared to $m$, as it reduces a factor $\mathcal{O}(m^3)$ to $\mathcal{O}(mp^2)$.

\section{Methods}
\label{sec:methods}
In this section, we will discuss aspects of the practical application of our spectral RFF method and also introduce the systems studied in the results section below. For the numerical examples, we will use the Gaussian RBF kernel or its periodic counterpart for periodic domains.

\subsection{Model Selection}

Before we can apply Algorithm~\ref{alg: rff_spectral}, we have to choose the number of features $p$, the truncation threshold for singular values, and as many hyper-parameters as required by the kernel function (e.g., the bandwidth for Gaussian RFF kernels). We select these parameters by cross-validating the VAMP score metric on randomly selected subsets of the data, as explained below. The only exception is the truncation threshold for singular values of $\M$. Here, we adopt the rule that all singular values smaller than $10^{-4}$ times the largest singular value are discarded. This choice is based on prior experience and works well in all considered examples.

The VAMP score metric is based on the \emph{Rayleigh variational principle}, which states that for reversible systems, the exact leading $K$ eigenfunctions are optimizers of the Rayleigh trace
\begin{equation*}
%\label{eq:vamp_variational}
    \sum_{i=0}^{K-1} \lambda_i(t) = \max_{\phi_0, \ldots, \phi_{K-1}} \sum_{i=0}^{K-1} \innerprod{\phi_i}{\cK^t  \phi_i}_\mu, \quad \text{s.t. } \innerprod{\phi_i}{\phi_j}_\mu = \delta_{ij}.
\end{equation*}
The sum to be maximized on the right-hand side is a special case of the slightly more general \emph{VAMP score} introduced for Koopman operators in~\cite{NOE2013, WU2020} and for the Koopman generator in~\cite{zhang2022}. When considering the generator, the maximum becomes a minimum instead. The variational principle also generalizes to non-reversible systems by considering singular functions instead of eigenfunctions~\cite{WU2020}. However, we only analyze reversible systems in what follows.
The VAMP score can be easily estimated upon replacing inner products by their empirical estimators, as shown in Section~\ref{subsubsec:finite_dim_approx}. The maximal VAMP score within a finite-dimensional model class can also be straightforwardly computed, see again~\cite{NOE2013}. In this way, we can simply compare the maximal VAMP score obtained for each choice of hyper-parameters, e.g., choice of kernel bandwidth $\sigma$ and number of Fourier features $p$, and choose the best.

To account for the estimation error stemming from the use of finite data, and also to avoid over-fitting, we employ a standard cross-validation approach, and first compute the optimal functions $\phi_0, \ldots, \phi_{K-1}$ for each model class on  a random subsample of the data (training data). We then recompute the VAMP score for the $K$-dimensional space spanned by $\phi_0, \ldots, \phi_{K-1}$ on the remaining data (test data). This process is repeated $n_{\mathrm{test}}$ times, where $n_{\mathrm{test}}$ is either 10 or 20 below, and we choose the model class optimizing the test score on average. The ratio of training data to test data is always chosen as a $75 / 25$ percent split.

\subsection{Spectral Analysis and Clustering}

Solution of the eigenvalue problem for the reduced matrix $\mathbf{R}$ provides estimates of the leading eigenvalues $\hat{\lambda}_i(t)$ or $\hat{\kappa}_i$, respectively, and their associated eigenfunctions $\hat{\psi}_i$. The first point to note is that~\eqref{eq:ev_problem_transposed} is a complex eigenvalue problem, so the eigenvectors can be scaled by arbitrary complex numbers. Before analyzing the eigenvectors, we simply apply a grid search over complex phase factors to determine the one that minimizes the imaginary part of all eigenvector entries and use the rescaled eigenvectors for further analysis.

Second, we transform the eigenvalues of the Koopman operator into \emph{implied timescales} defined as
\begin{equation*}
    t_i = -\frac{t}{\log(\hat{\lambda}_i(t))},
\end{equation*}
which have units of time, and can be interpreted as physical relaxation timescales for the $i$-th eigenmode of the Koopman operator. Since timescales are independent of the lag time in the absence of projection and estimation error~\cite{Sarich2010}, we monitor convergence of implied timescales with increasing lag time $t$ in order to select an optimal lag time~\cite{Prinz2011c}. We also compute the timescales when approximating the generator, in this case they are simply given by
\begin{equation*}
    t_i = -\frac{1}{\hat{\kappa}_i}.
\end{equation*}
In order to analyze metastable sets, we evaluate the top eigenfunctions $\hat{\psi}_i$ at all data points, and apply the spectral clustering method PCCA~\cite{DEUFLHARD2005} to assign each data point $x_l$ to each metastable set $q$ with a degree of membership $\chi_q(x_l)$, which is a number between zero and one. We then identify data point $x_l$ as part of metastable set $q$ if $\chi_q(x_l) \geq 0.6$, the remaining data points are treated as transition states.

\subsection{Systems and Simulation Setups}
We study four different examples in order to illustrate different aspects of the performance of the proposed method. The first is overdamped Langevin dynamics~\eqref{eq:od_langevin} in a two-dimensional model potential, which is a minor modification of the \emph{Lemon-Slice potential}~\cite{Bittracher2018}, given in polar coordinates as
\begin{equation}
\label{eq:lemonslice}
    V(r, \varphi) = \cos(4\varphi) + 10(r - 1)^2 + 1 +\frac{1}{r} + \frac{1}{\cos \varphi},
\end{equation}
see Figure~\ref{fig:lemon_slice_intro} for a visualization. We can easily generate large amounts of simulation data by applying the standard Euler--Maruyama integrator, the elementary simulation time step is set to $\Delta_t = 10^{-3}$. Friction $\gamma$ and inverse temperature $\beta$ are both set to one. Reference implied timescales are obtained from a Markov state model based on a $k$-means discretization of the two-dimensional state space, using 50 discrete states and $m = 10^5$ data points.

The second system is molecular dynamics simulation data of the alanine dipeptide in explicit water, at temperature $T = 300\,\mathrm{K}$, employing the Langevin thermostat. We generated a total of one million data points at $1\,\mathrm{ps}$ time spacing. Reference implied timescales were computed by a Markov state model in the well-known two-dimensional space of backbone dihedral angles $\phi$ and $\psi$, using a $30\times30$ box discretization and all one million data points.

The third example is molecular dynamics simulation data of the deca alanine peptide in explicit water at temperature $T = 300\,\mathrm{K}$, see \cite{NuesSchn16} for a description of the simulation setup. The data comprises a total of four million time steps at $1 \,\mathrm{ps}$ time spacing. As reference values, we refer to the MSM analysis presented in \cite{Nueske2021}, which is built on a 500-state $k$-means discretization after applying time-lagged independent component analysis (TICA)~\cite{PEREZ2013} in the space of the peptide's sixteen backbone dihedral angles.

Finally, we re-analyze the 39-residue protein NTL9, which was simulated on the Anton supercomputer by D.E. Shaw Research, see \cite{LINDORFF2011} for details. The data comprises around 1.2 million frames corresponding to a time spacing of $\Delta_t = 2\, \mathrm{ns}$. As a reference, we apply linear TICA on the set of 666 minimal heavy atom inter-residue distances. We also rank these distance according to the fraction of simulation time where a contact is formed, that is, the distance does not exceed $0.35\,\mathrm{nm}$, see~\cite{Nueske2021}, and calculate TICA models using only the first $[20, 100, 200, 300, 400, 500, 666]$ of these distances.

\section{Results}
\label{sec:results}

\subsection{Lemon-Slice Potential}

\begin{figure}
    \centering
    \includegraphics[width=0.48\textwidth]{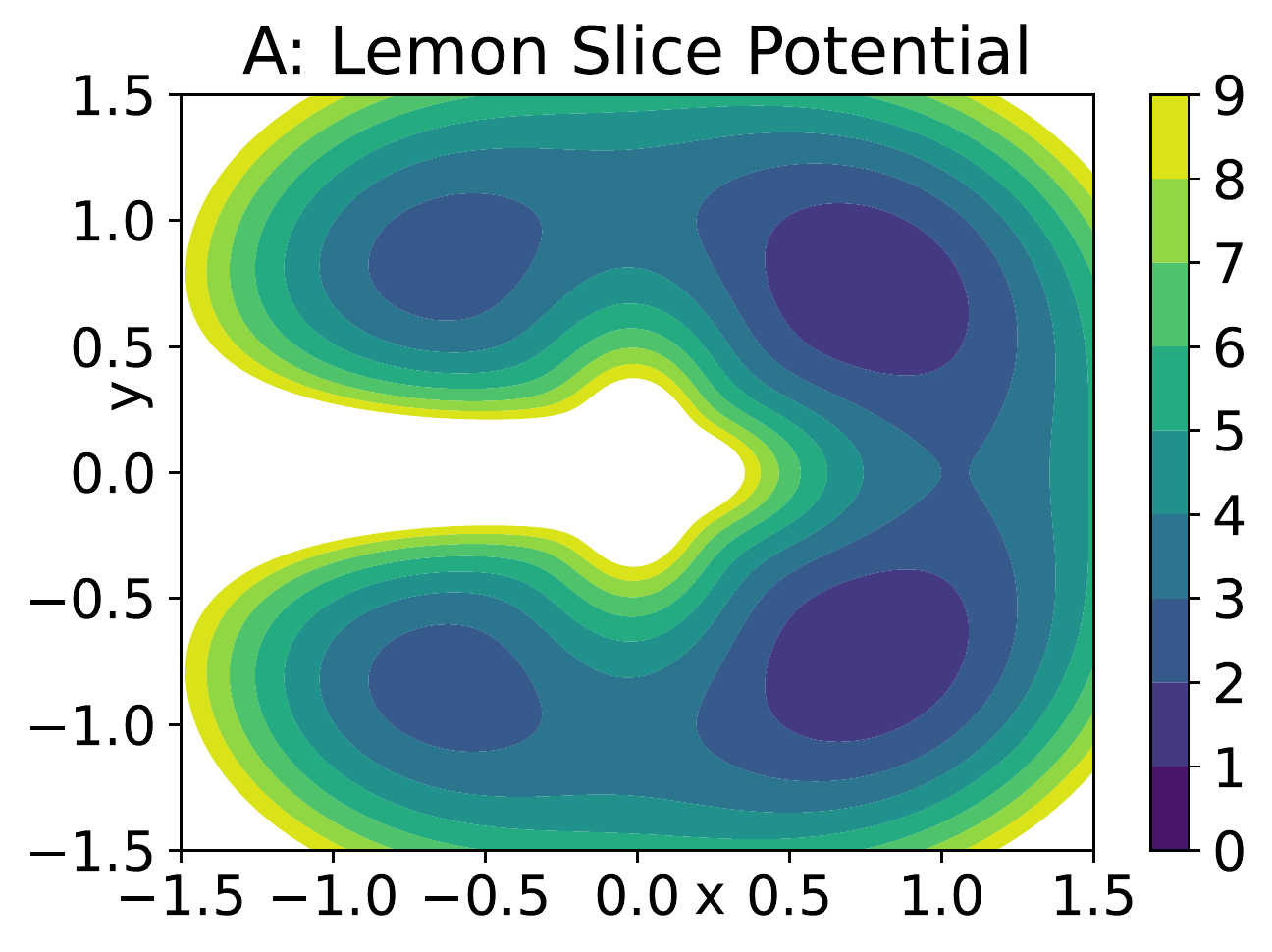}
    \includegraphics[width=0.48\textwidth]{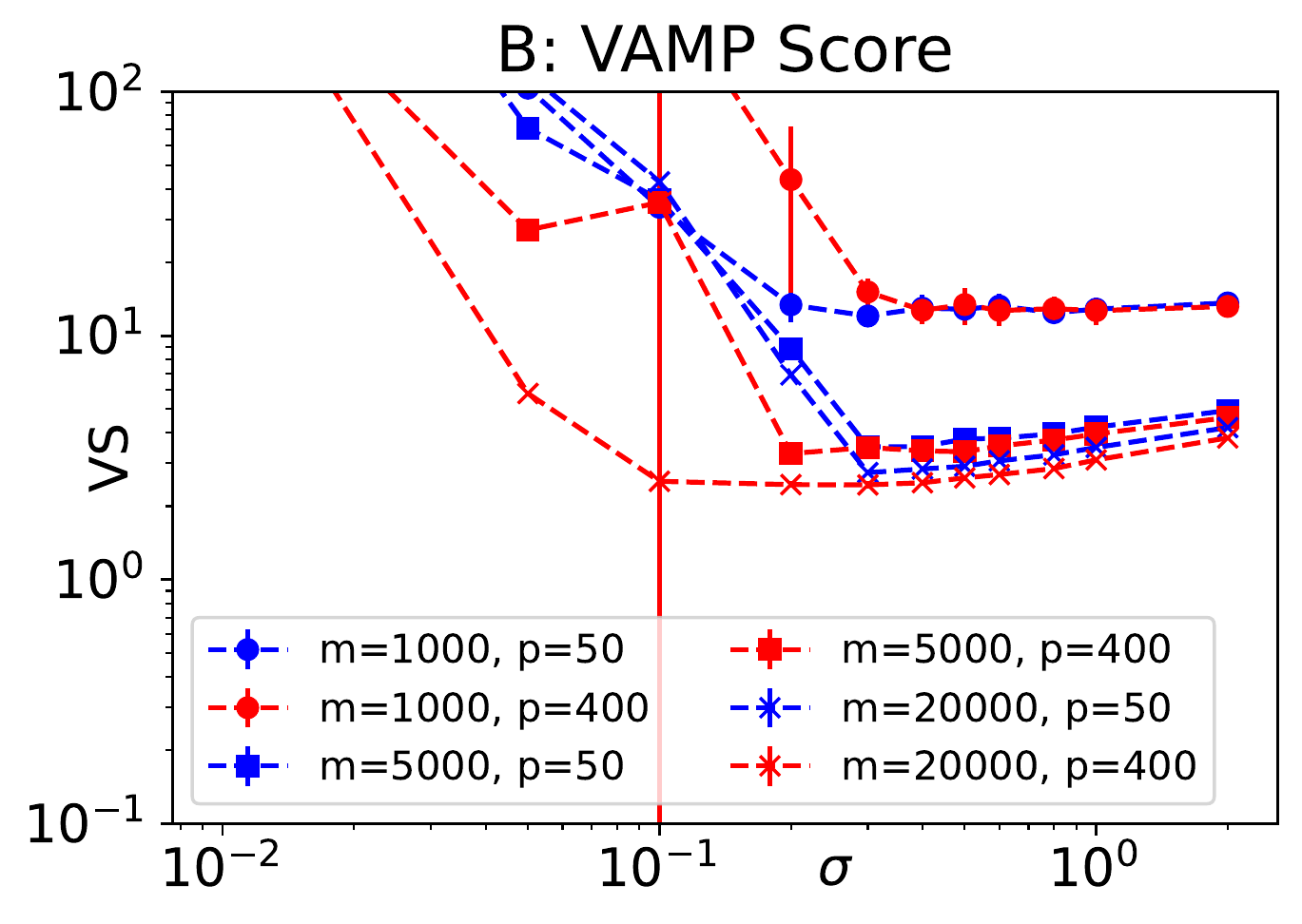}
    \includegraphics[width=0.48\textwidth]{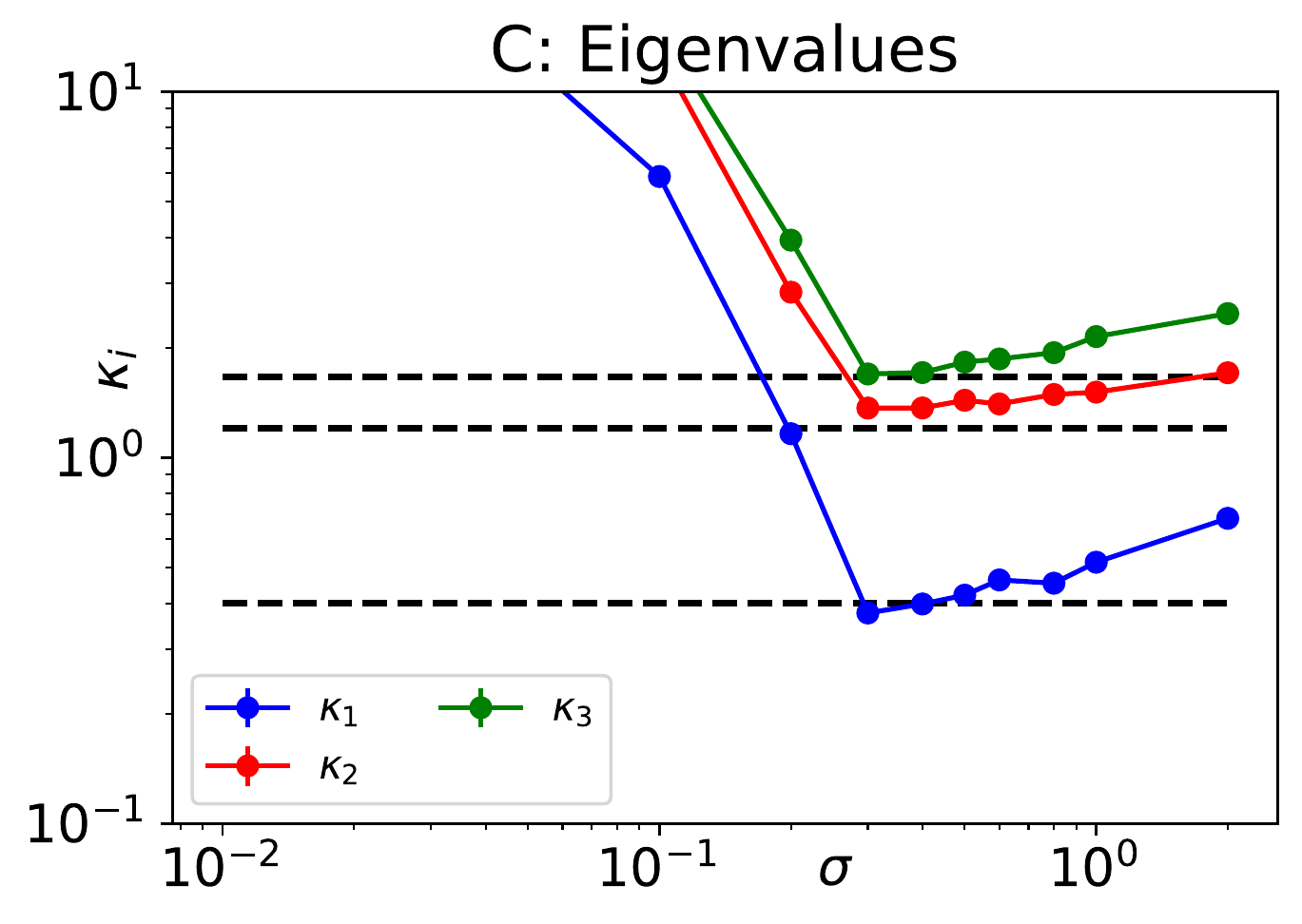}
    \includegraphics[width=0.48\textwidth]{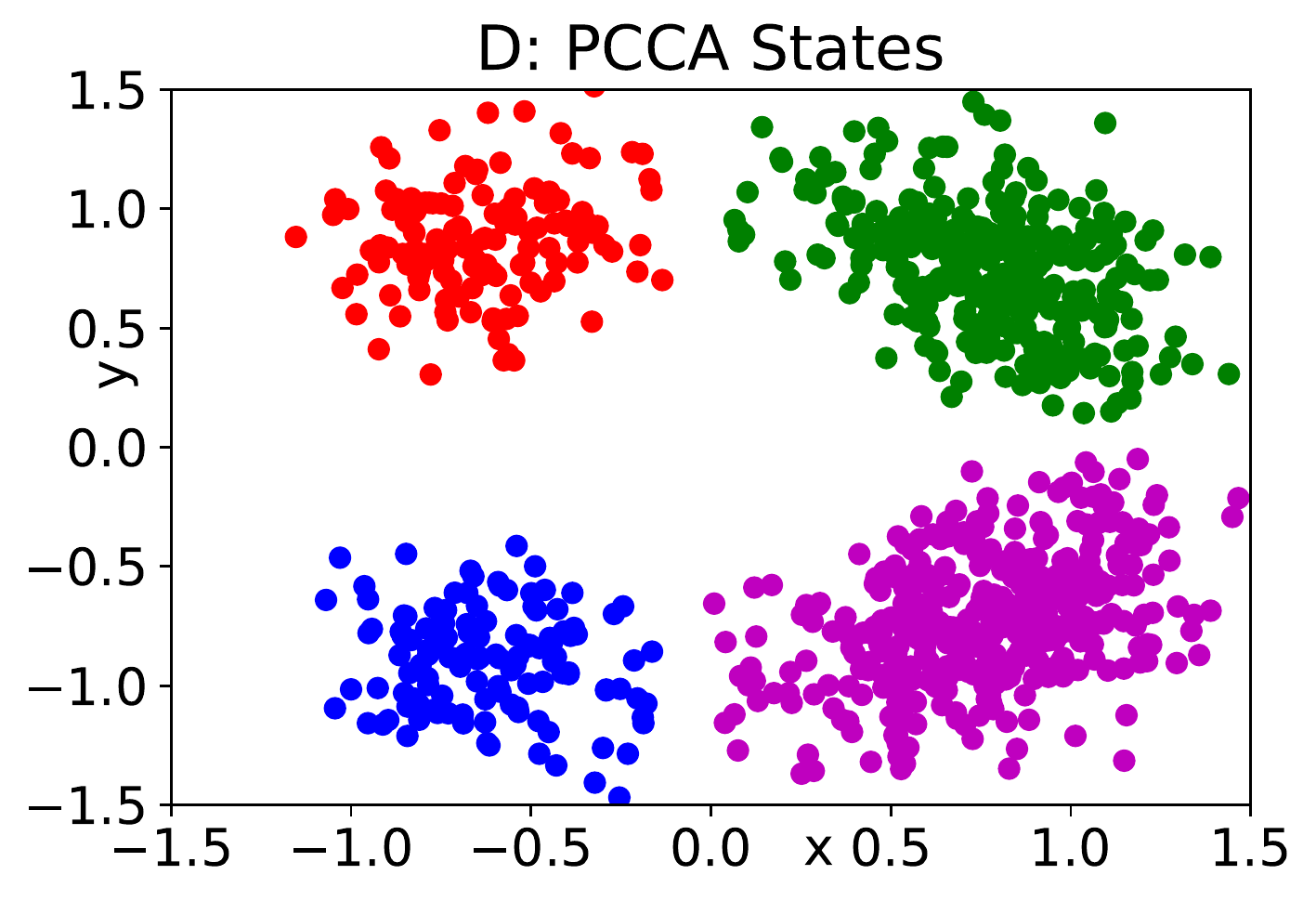}
    \caption{Results for the Lemon-Slice potential. A: Contour of the potential~\eqref{eq:lemonslice}. B: VAMP score as a function of the kernel bandwidth for different data sizes $m$ and feature numbers $p$. C: Leading non-trivial eigenvalues of the generator for $m = 5000$ and $p = 50$, as a function of the bandwidth. Black lines indicate the Markov model reference values. D: Decomposition into four metastable states based on eigenvectors for $\sigma = 0.4$, $p = 50$, $m = 5000$. All error bars are based on twenty independent simulations.}
    \label{fig:lemon_slice_intro}
\end{figure}

We start by analyzing the Lemon-Slice potential~\eqref{eq:lemonslice}, shown in Figure~\ref{fig:lemon_slice_intro}~A, which features four metastable states corresponding to the wells of the potential energy. We use this system as a first test case for approximating the generator $\cL$, due to the availability of all relevant quantities in analytical form.

Figure~\ref{fig:lemon_slice_intro}~B shows the VAMP score as a function of the bandwidth for a selection of data sizes $m$ and feature numbers $p$. We see that while $m = 1000$ data points are not sufficient, for $m = 5000$ data points, the score stabilizes for a range of bandwidths $\sigma \in [0.3, 0.8]$, and already attains optimal values for very small $p$. We confirm this finding by plotting the first three non-trivial eigenvalues of the generator, for $m = 5000$ and $p = 50$, as a function of the bandwidth in Figure~\ref{fig:lemon_slice_intro}~C. Indeed, accurate estimates for eigenvalues can be obtained in the suggested regime of bandwidths. After calculating the eigenvectors of the reduced matrix for $\sigma = 0.4$, and applying the PCCA analysis, we see that the metastable structure is perfectly recovered, see Figure~\ref{fig:lemon_slice_intro}~D. Thus, the VAMP score reliably guides the hyper-parameter search towards a very efficient approximation of the leading spectral components of this system.

\subsection{Alanine Dipeptide}

\begin{figure}
    \centering
    \includegraphics[width=0.48\textwidth]{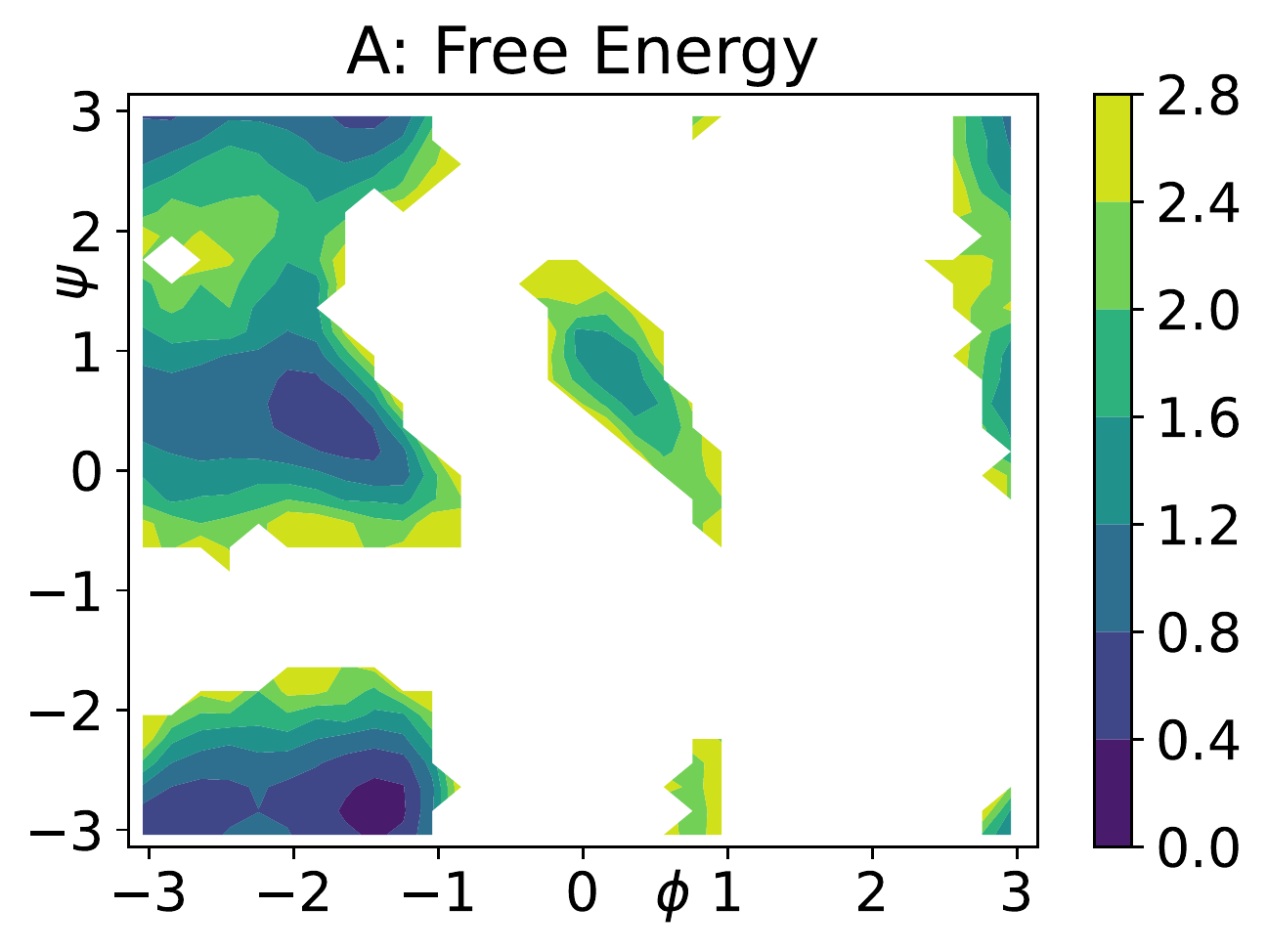}
    \includegraphics[width=0.48\textwidth]{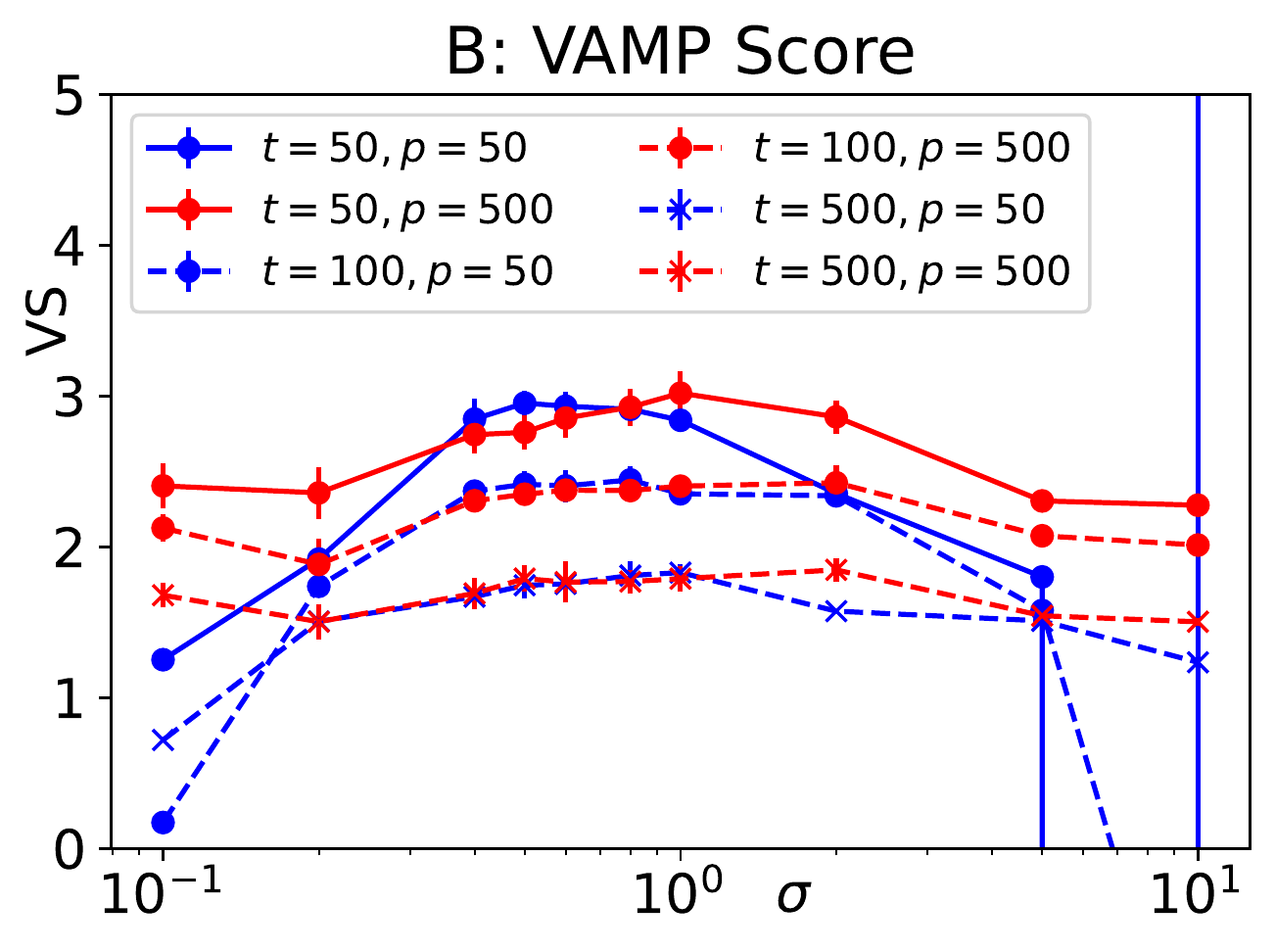}
    \includegraphics[width=0.48\textwidth]{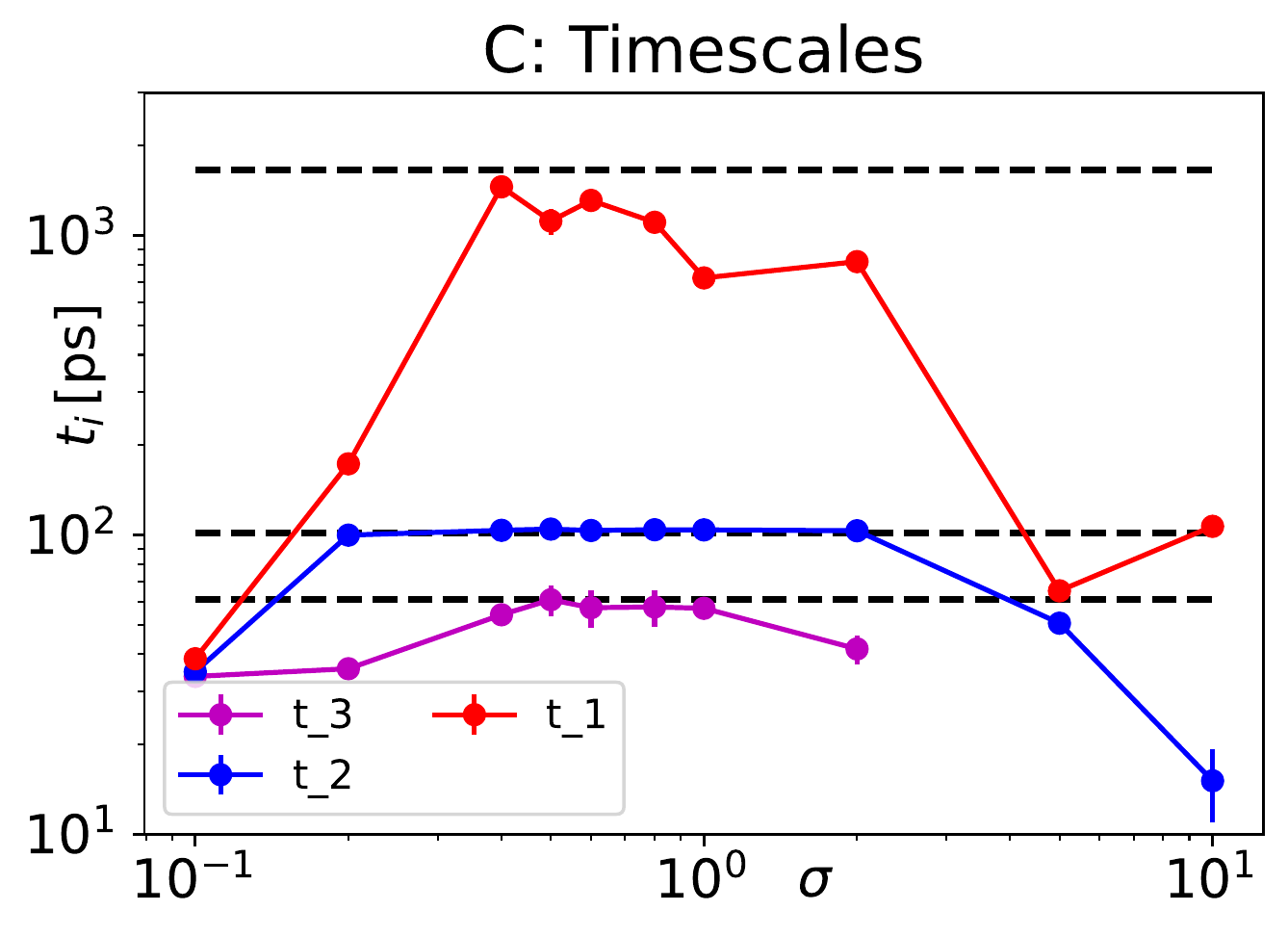}
    \includegraphics[width=0.48\textwidth]{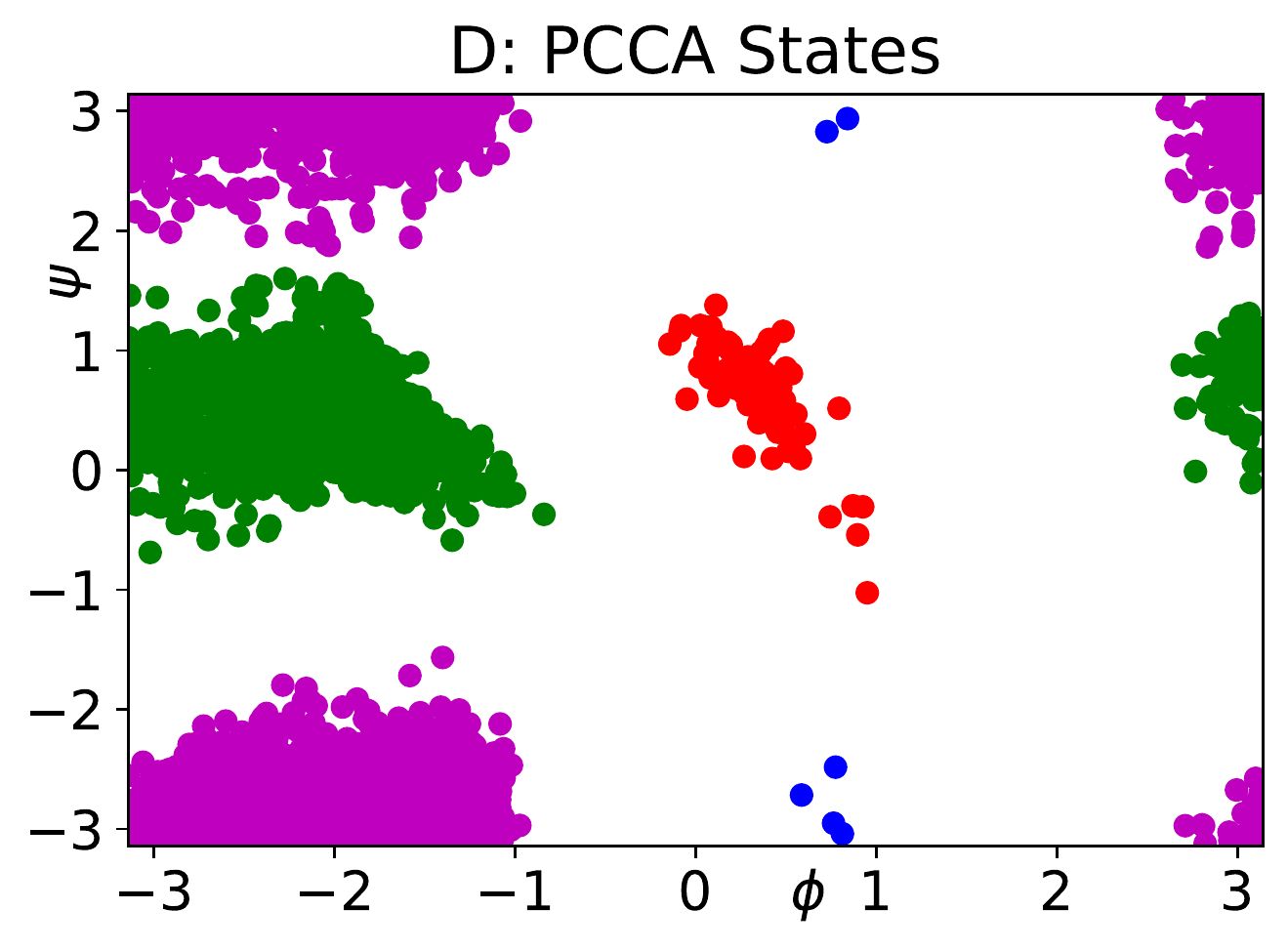}
    \caption{Results for alanine dipeptide. A: Free Energy (in $\mathrm{kJ/mol}$) in two-dimensional dihedral space. B: VAMP Score for selected feature sizes $p$ and lag times $t$ as a function of the kernel bandwidth. C: Implied timescales for $t = 100\,\mathrm{ps}$ and $p = 50$ as a function of the bandwidth. D: Metastable decomposition obtained for $\sigma = 0.6,\, p = 50$ and $t = 100\,\mathrm{ps}$.}
    \label{fig:ala2}
\end{figure}

For alanine dipeptide, we approximate the Koopman operator with lag time $t$, using the periodic Gaussian kernel on the well-known two-dimensional reaction coordinate space given by the dihedral angles $\phi$ and $\psi$. Figure~\ref{fig:ala2}~A shows the familiar free energy landscape of the system in the reaction coordinate space. For all RFF-based approximations, we downsample the data set to $m = 20\ts000$ points, corresponding to a time spacing of $50\,\mathrm{ps}$. We apply the same model selection protocol as before, and find that a range of kernel bandwidths $\sigma \in [0.4, 1.0]$ can be stably identified as optimal in terms of the VAMP score. We observe in Figure~\ref{fig:ala2}~B that again a small number of Fourier features $p = 50$ is sufficient to arrive at an optimal VAMP score. We also confirm that the optimal parameter regime is stable across different lag times. Note that the VAMP score necessarily decreases with increasing lag time, as all eigenvalues but $\lambda_0(t)$ decay exponentially with $t$.
As shown in Figure~\ref{fig:ala2}~C, the optimal range of bandwidths indeed allows for accurate estimation of the three leading implied timescales of the system. Further, we also verify that the metastable decomposition of the dihedral space into four states is correctly recovered using the RFF model, see Figure~\ref{fig:ala2}~D. The dramatic rank reduction for the kernel matrix achieved by using a small number of Fourier features allows us to easily compute eigenvalues and test scores for a broad range of parameters, using $m = 20\ts000$ data points in real space. Solving the same problem by means of standard methods would be significantly more costly if the same data size was used.

\subsection{Deca Alanine}
We use the deca alanine example to demonstrate that, by means of low-rank kernel methods, generator models can be efficiently learned on a reaction coordinate space that is more than just one- or two-dimensional, and that dynamical hypotheses can be tested this way. We project the system onto the space of its sixteen interior backbone torsion angles. The generator of the full molecular dynamics is not readily available in closed form. It is well-known, however, that if observed only on a subset of position space, many systems driven by thermostatted molecular dynamics behave like the reversible overdamped Langevin dynamics at long timescales~\cite{Stoltz2010}. For that reason, we set the full space diffusion matrix to $\frac{1}{\beta}$ times the identity and employ our reversible estimator. The effective diffusion~\eqref{eq:effective_diffusion} then turns into
\begin{align*}
    a^\xi(z) = \frac{1}{\beta}\nabla \xi^\top(x) \nabla \xi(x),
\end{align*}
where $\nabla \xi$ is the Jacobian of the mapping from Euclidean coordinates to the peptide's backbone dihedral angles, which can be evaluated analytically. Note that this approach can be seen as a dynamical hypothesis: we choose to model the position space dynamics as an overdamped Langevin process, and test if its projection onto the dihedral angle space can recover all relevant features of the full MD simulation.

\begin{figure}
    \centering
    \includegraphics[width=0.49\textwidth]{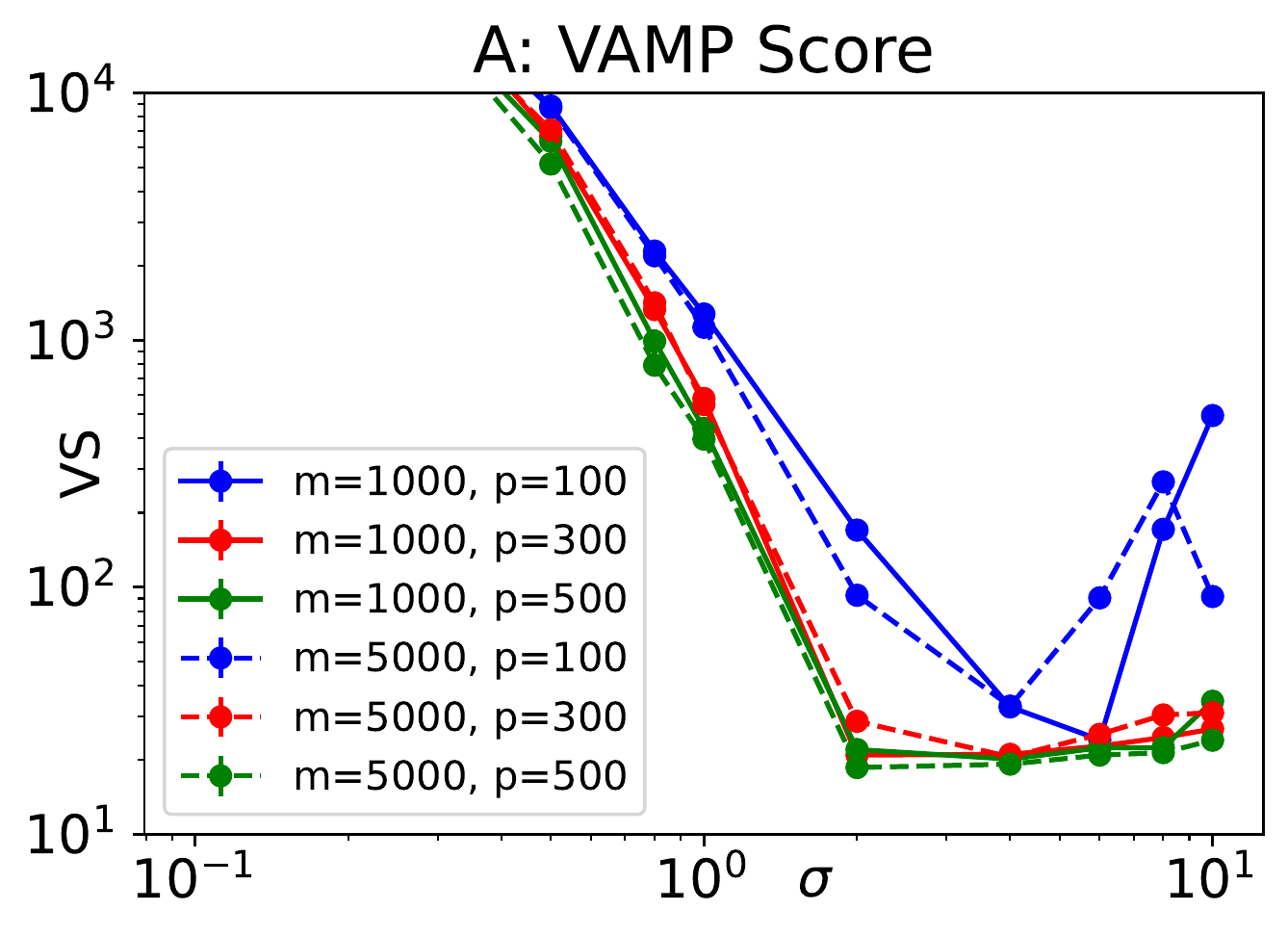}
    \includegraphics[width=0.49\textwidth]{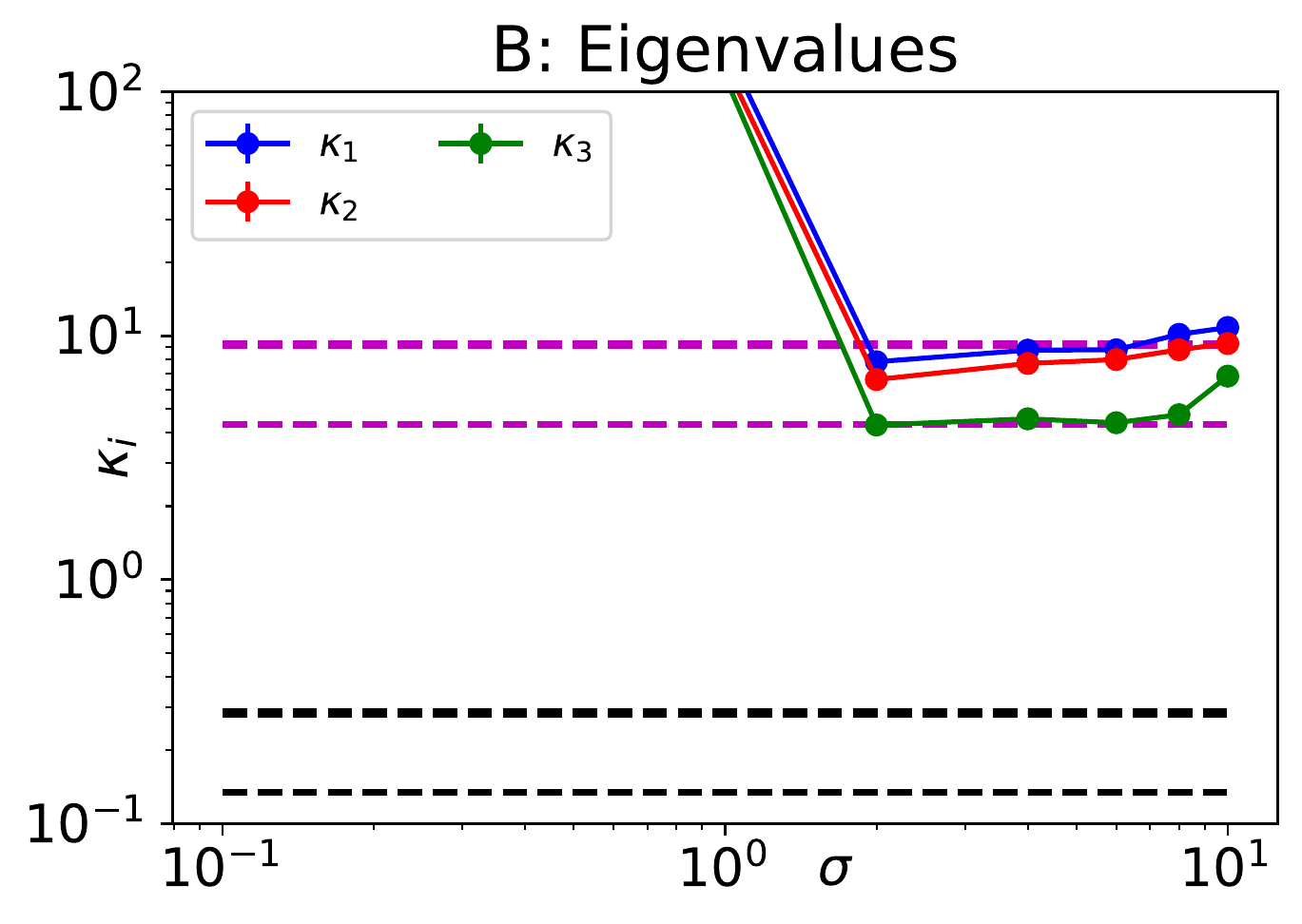} \\[1ex]
    \begin{minipage}[t]{0.2\textwidth}
        \centering C: \\
        \includegraphics[width=\textwidth]{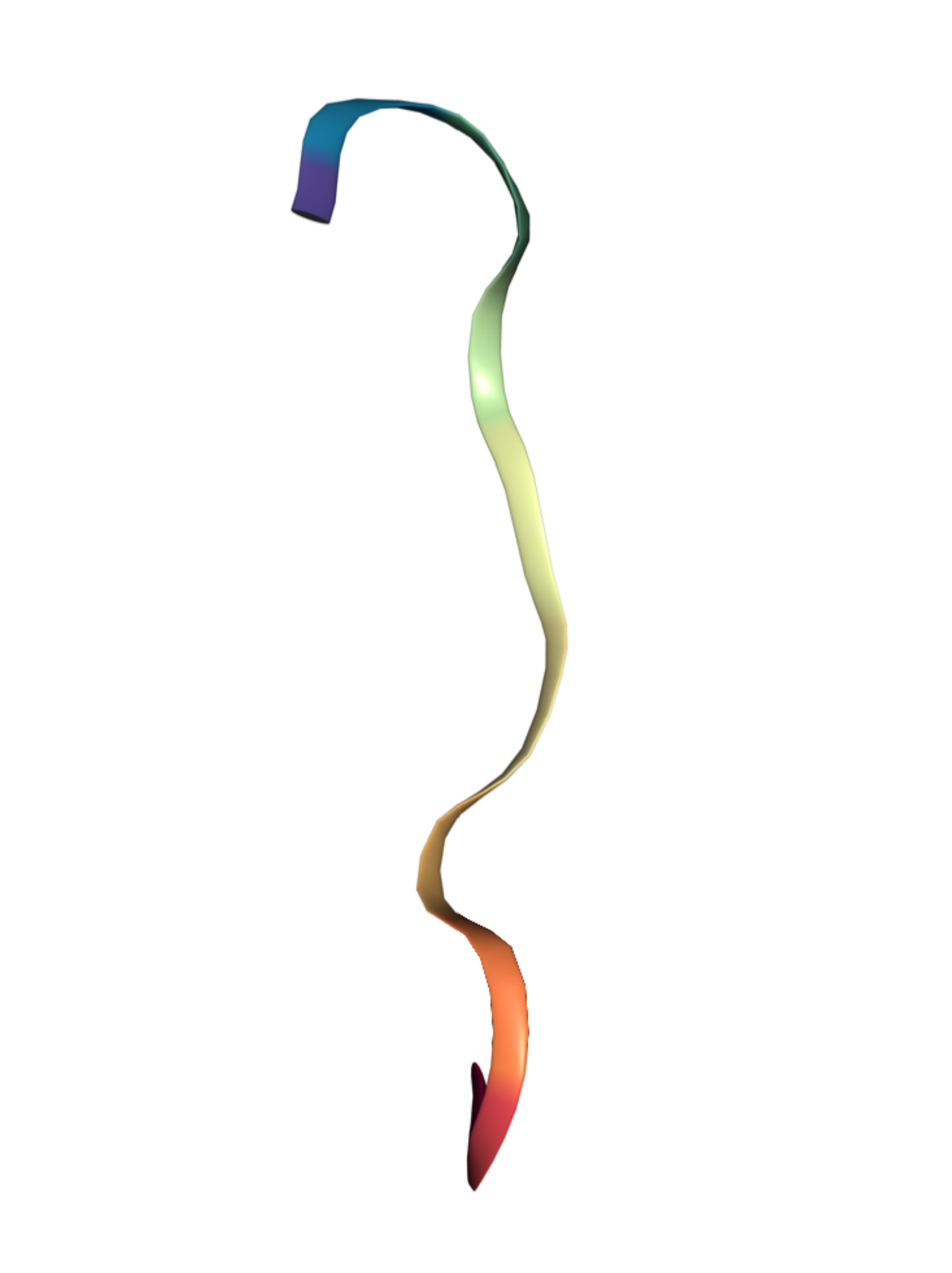}
    \end{minipage}
    \begin{minipage}[t]{0.2\textwidth}
        \centering D: \\
        \includegraphics[width=\textwidth]{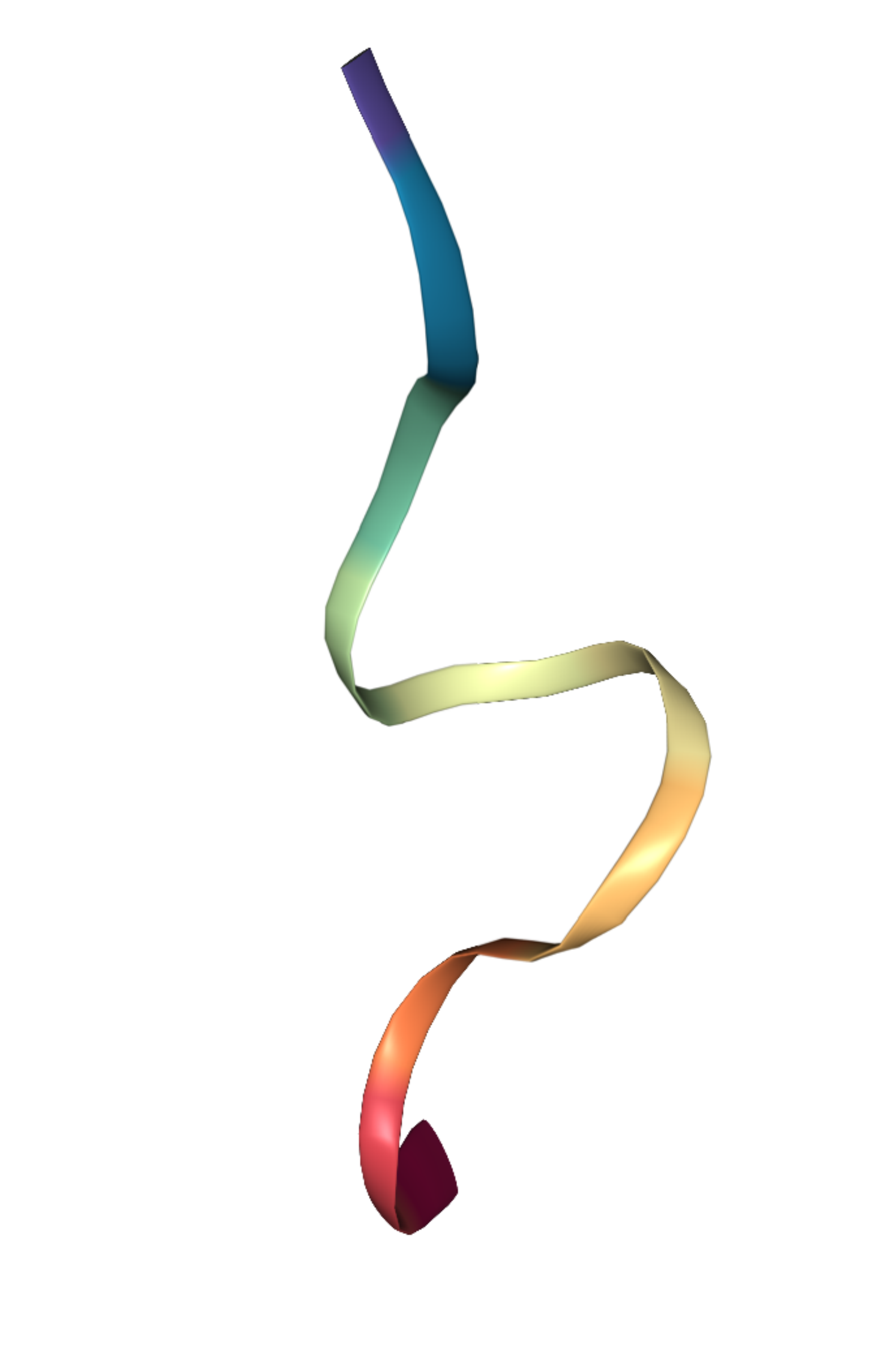}
    \end{minipage}
    \begin{minipage}[t]{0.2\textwidth}
        \centering E: \\
        \includegraphics[width=\textwidth]{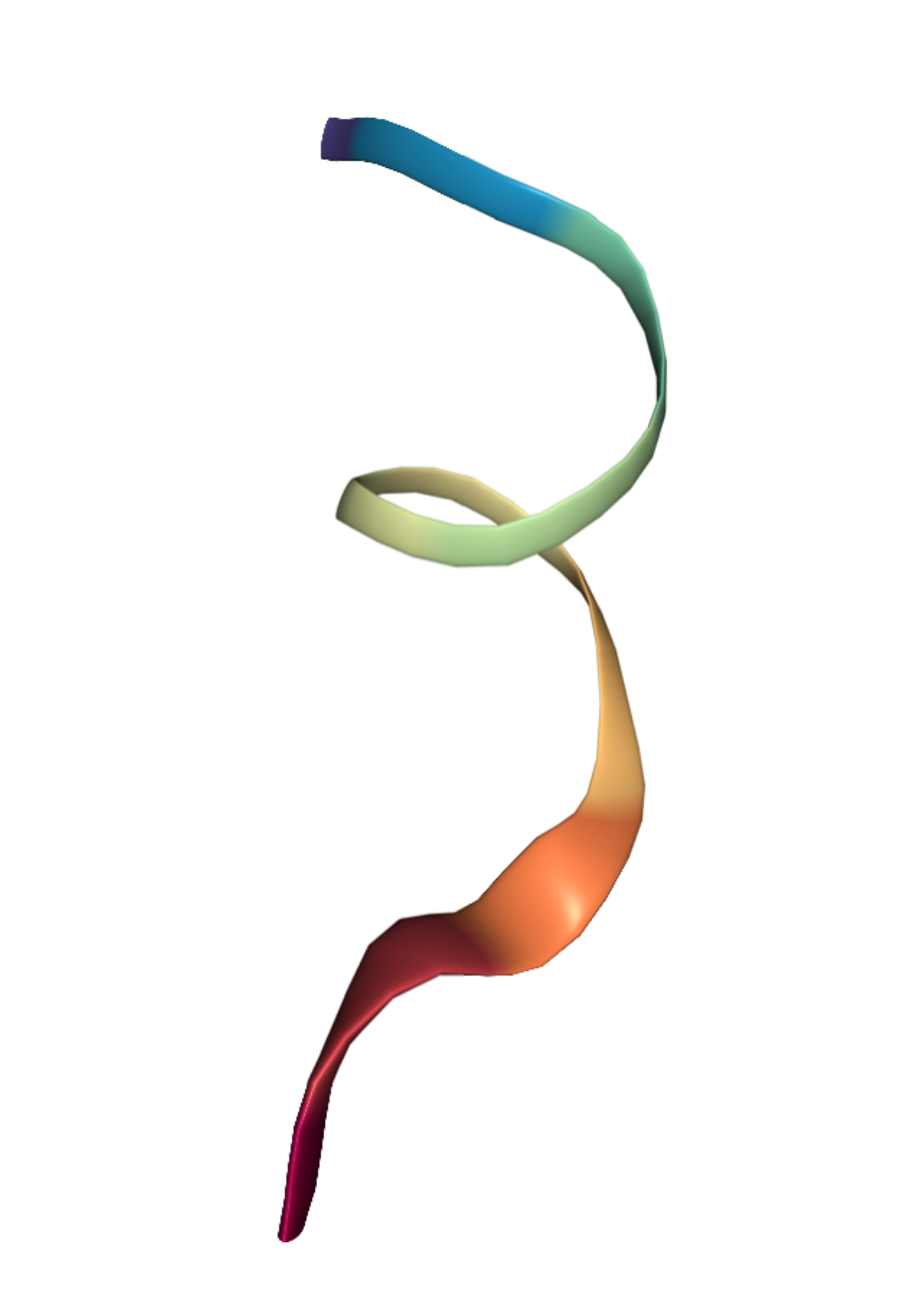}
    \end{minipage}
    \begin{minipage}[t]{0.2\textwidth}
        \centering F: \\
        \includegraphics[width=0.85\textwidth]{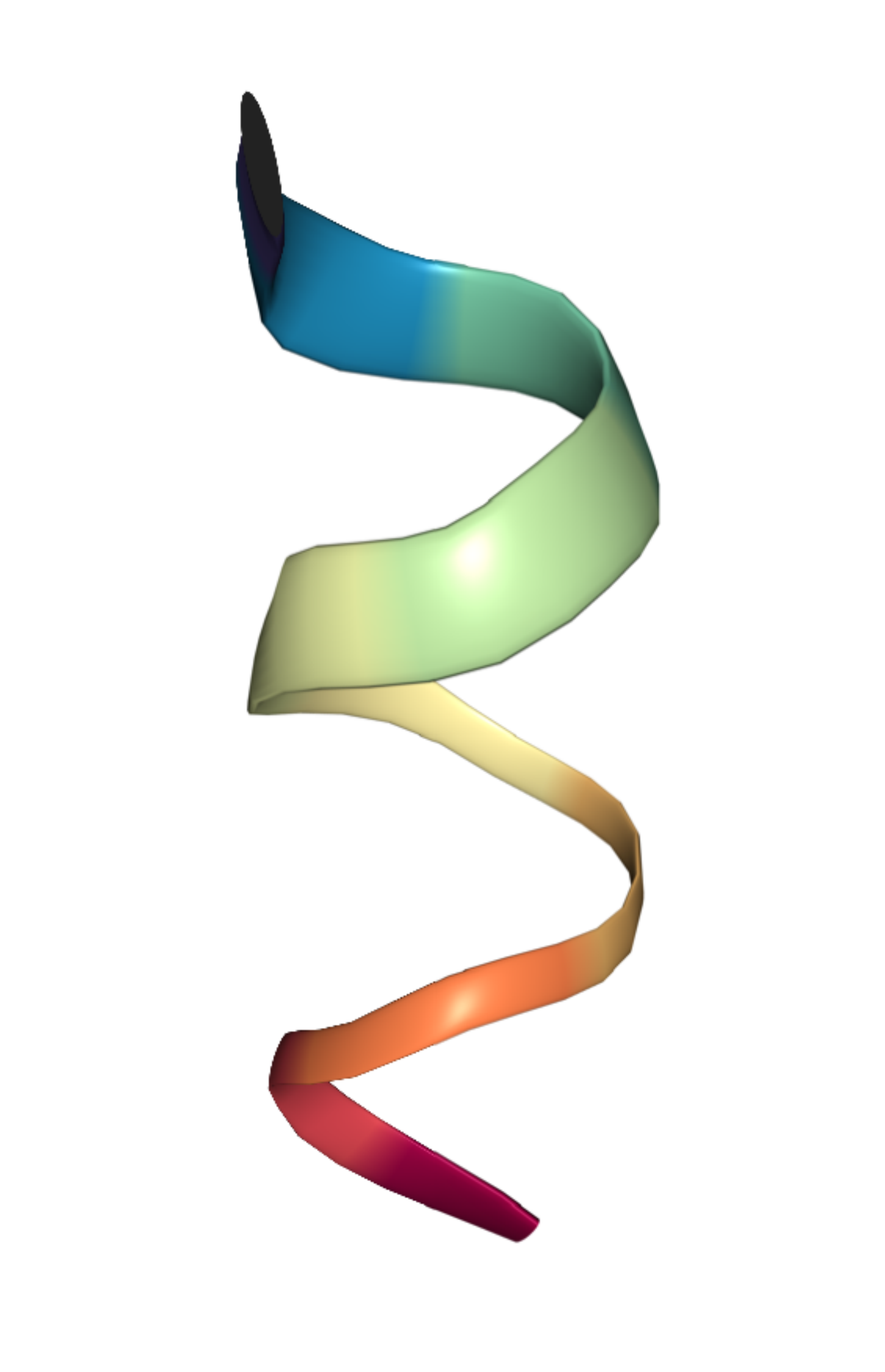}
    \end{minipage}
    \vspace*{-2ex}
    \caption{Results for the Koopman generator on backbone dihedral angle space of the deca alanine peptide. A: VAMP score for selected data sizes $m$ and feature sizes $p$ as a function of the kernel bandwidth. B: Eigenvalues for $m = 1000$ and $p =300$ as a function of the bandwidth. The reference MSM results are shown in black. Re-scaling MSM eigenvalues by the average ratio between optimal RFF and MSM eigenvalues leads to the magenta lines. C--F: Representative structures for each of the four PCCA states based on the RFF model at $m =1000$, $p = 300$, $\sigma = 4.0$.}
    \label{fig:ala10_spectrum}
\end{figure}

We find in Figure~\ref{fig:ala10_spectrum}~A that a kernel bandwidth $\sigma \in [2.0, 8.0]$ leads to an optimal VAMP score for the projected generator. Using about $m = 1000$ data points in real space and $p = 300$ Fourier features is sufficient to obtain stable results, rendering the calculation highly efficient. The optimality of this regime of bandwidths in confirmed by plotting the first three non-trivial eigenvalues of the generator, as shown in Figure~\ref{fig:ala10_spectrum}~B. We note that these eigenvalues differ from those of the reference MSM by a constant factor. This is due to the re-scaling of time inherent in assuming overdamped Langevin dynamics in position space, rendering the effective dynamics too fast. However, we also see in Figure~\ref{fig:ala10_spectrum} that all slow eigenvalues are indeed re-scaled by the same factor. Moreover, the PCCA analysis of the slowest eigenvectors obtained for the optimal RFF model also recovers the metastable states of the deca alanine system, corresponding to the folded and unfolded states as well as two intermediates. The effective overdamped Langevin model learned by the kernel method thus retains all major features of the original molecular dynamics, up to the re-scaling of time.

\subsection{NTL9}

\begin{figure}
    \centering
    \includegraphics[width=0.48\textwidth]{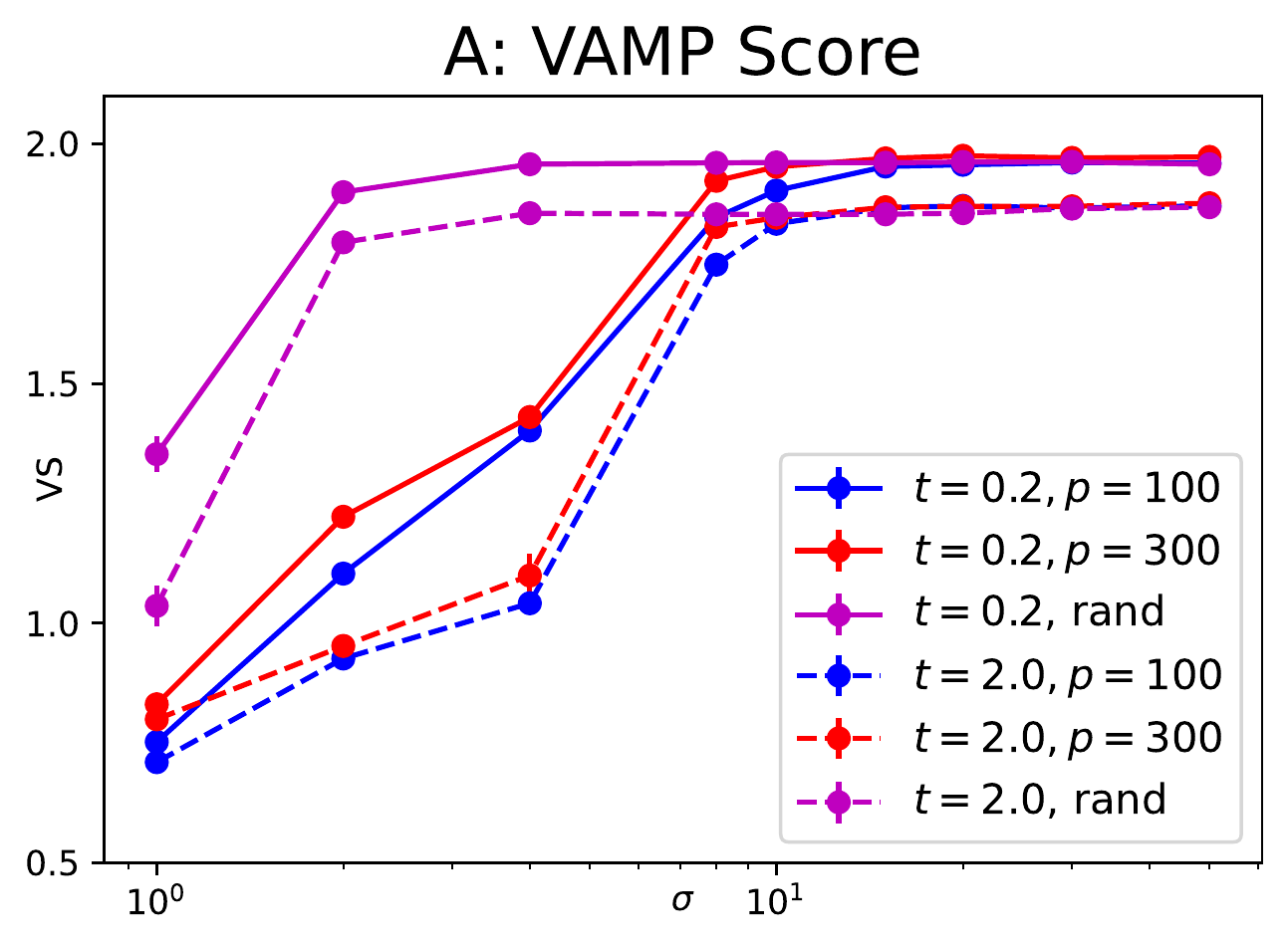}
    \includegraphics[width=0.48\textwidth]{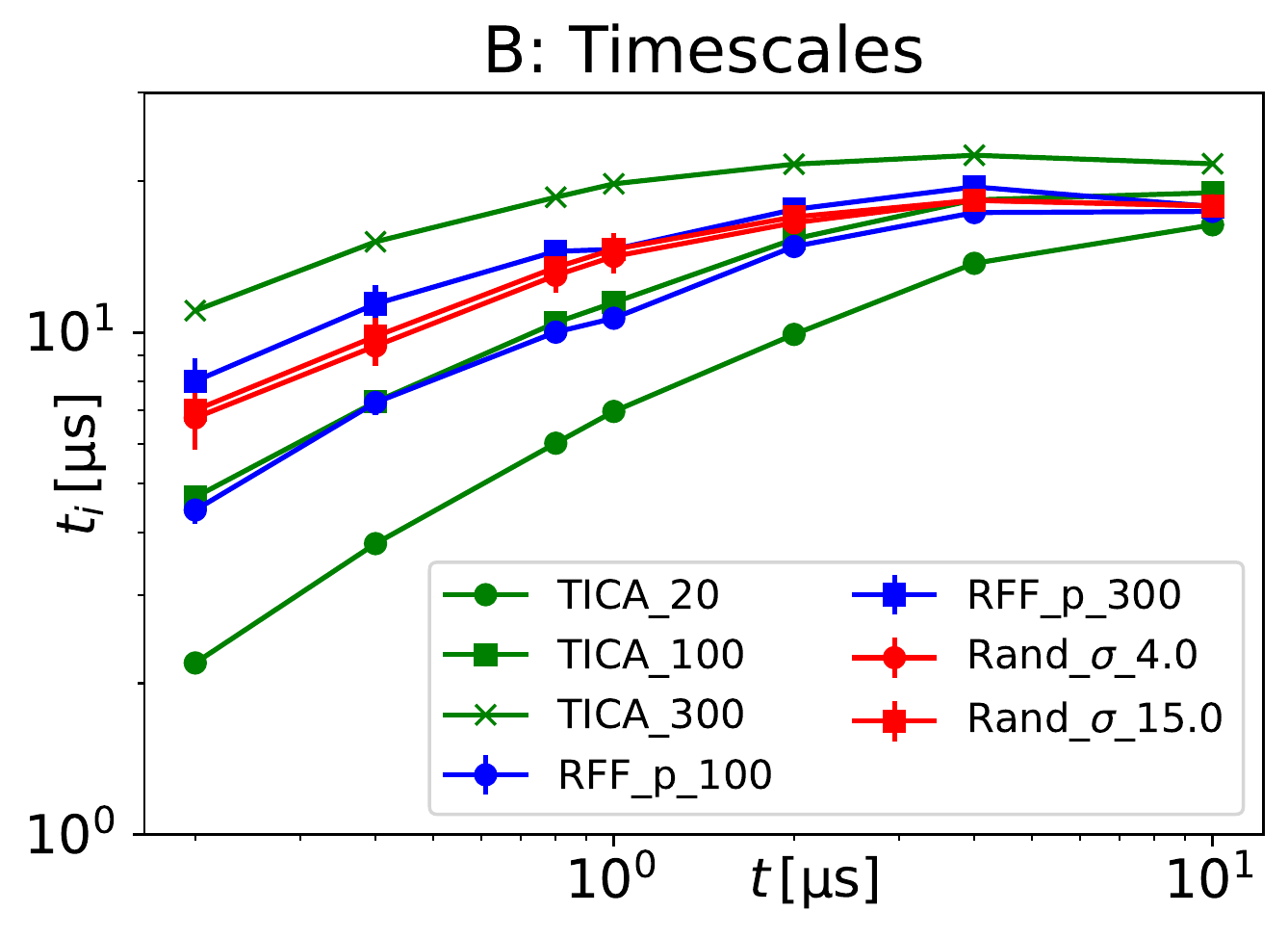}
    \caption{Results for NTL9 protein. A: VAMP score for Gaussian RFF approximation as a function of the kernel bandwidth $\sigma$. Red and blue lines show the results using all 666 distances, for different lag times and different numbers of Fourier features $p$. The magenta lines show the average values over ten random selections of only $50$ distances, with $p = 300$ fixed. B: Slowest implied timescale $t_1$ as a function of the lag time $t$. Green lines show estimates based on linear TICA using the top-ranked 20, 100, and 300 distances. Blue lines show RFF-based estimates on all distances using different values of $p$, using $\sigma = 15.0$. Red lines are averages over the above-mentioned random subsamples of the distance coordinates, where $p = 300$ is fixed.}
    \label{fig:ntl9}
\end{figure}

We use the NTL9 data set to demonstrate once again the ability of the proposed method to efficiently and robustly analyze large data sets. We downsample the available trajectory data to around $m = 15.000$ data points at $200\,\mathrm{ns}$ time spacing. As mentioned before, $m = 15.000$ is still a significant data size for kernel methods. We project the MD simulations onto the 666-dimensional space of minimal heavy atom inter-residue distances. It is known from previous publications that these coordinates capture the folding process of NTL9 very well~\cite{Boninsegna2018,Nueske2021}. Indeed, just using linear TICA on the top-ranked 300 distances allows for an accurate estimation of the slowest implied timescale $t_1$, associated to the folding process, see the top green line in Figure~\ref{fig:ntl9}~B.

Figure~\ref{fig:ntl9}~A shows that using Gaussian bandwidth parameters $\sigma \in [10, 50]$ can be identified as optimal in terms of the VAMP score, again across different lag times. In Figure~\ref{fig:ntl9}~B, we first compare the slowest timescale $t_1$ obtained for $\sigma = 15$ as a function of the lag time, for different feature numbers $p$ (blue lines). We see that already for $p \geq 300$, satisfactory convergence of $t_1$ with $t$ can be obtained, reducing the computational cost by about a factor two thousand compared to the full kernel eigenvalue problem. Interestingly, timescale convergence improves with $p$ in about the same way as the performance of TICA improves as more and more of the ranked distances are added. However, we show that the kernel method is more robust with respect to the selection of input coordinates, by randomly selecting $d = 50$ out of the 666 distances, and recomputing both the VAMP score and implied timescales for different bandwidths, keeping the number of Fourier features fixed at $p = 300$. We see in Figures~\ref{fig:ntl9}~A--B that both the optimal value of the VAMP score as well as the timescale estimates remain stable across random selections of the input distances (magenta lines in panel A, red lines in panel B). Convergence of timescales is therefore mainly dependent on the number of Fourier features $p$, which can be automatically tuned using the VAMP score, and not on the selection of input coordinates. Moreover, we also find that under the random distance selection, the optimal regime of kernel bandwidths changes as the dimension of the input coordinate space is significantly smaller. This is reliably reflected by the behavior of the VAMP score, see again Figure~\ref{fig:ntl9}~A.

\begin{figure}
\centering
    \includegraphics[width=0.48\textwidth]{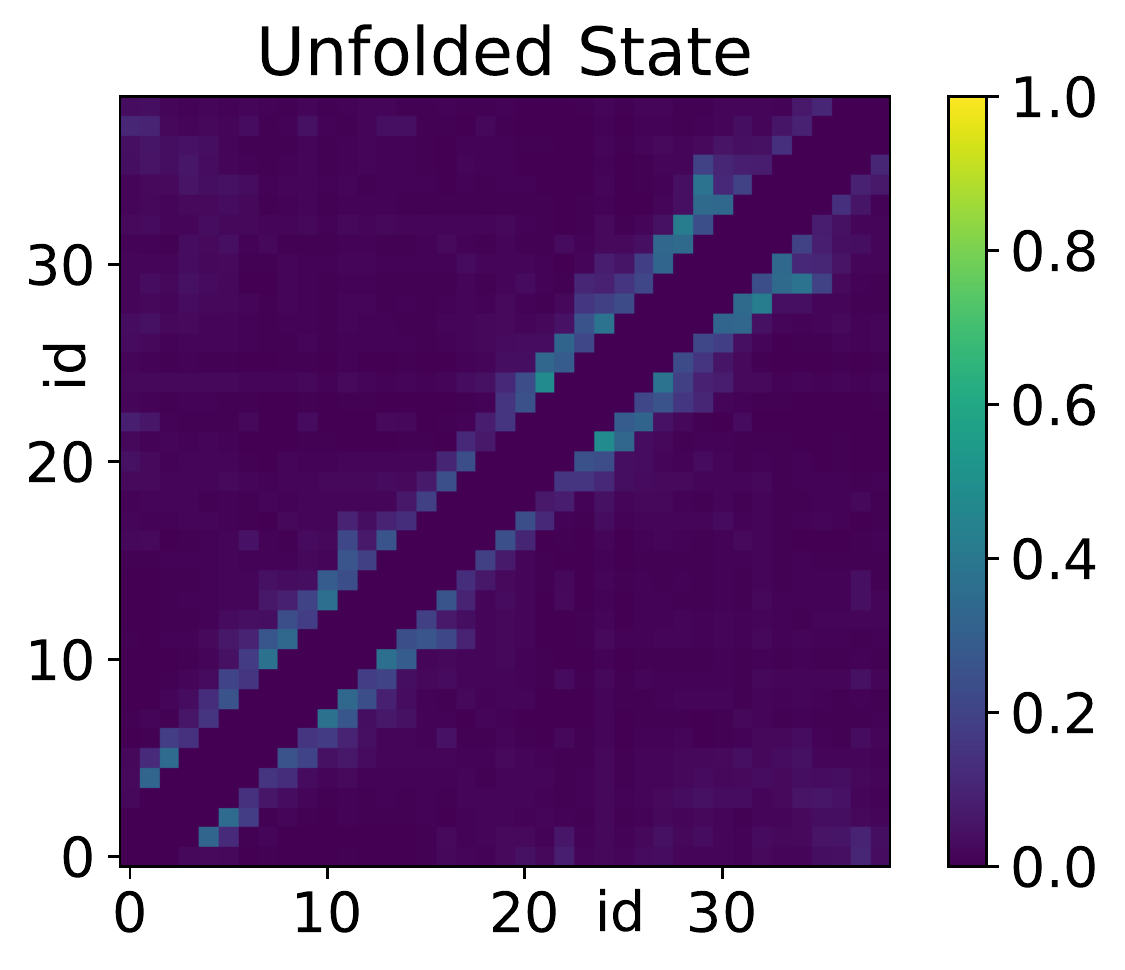}
    \includegraphics[width=0.48\textwidth]{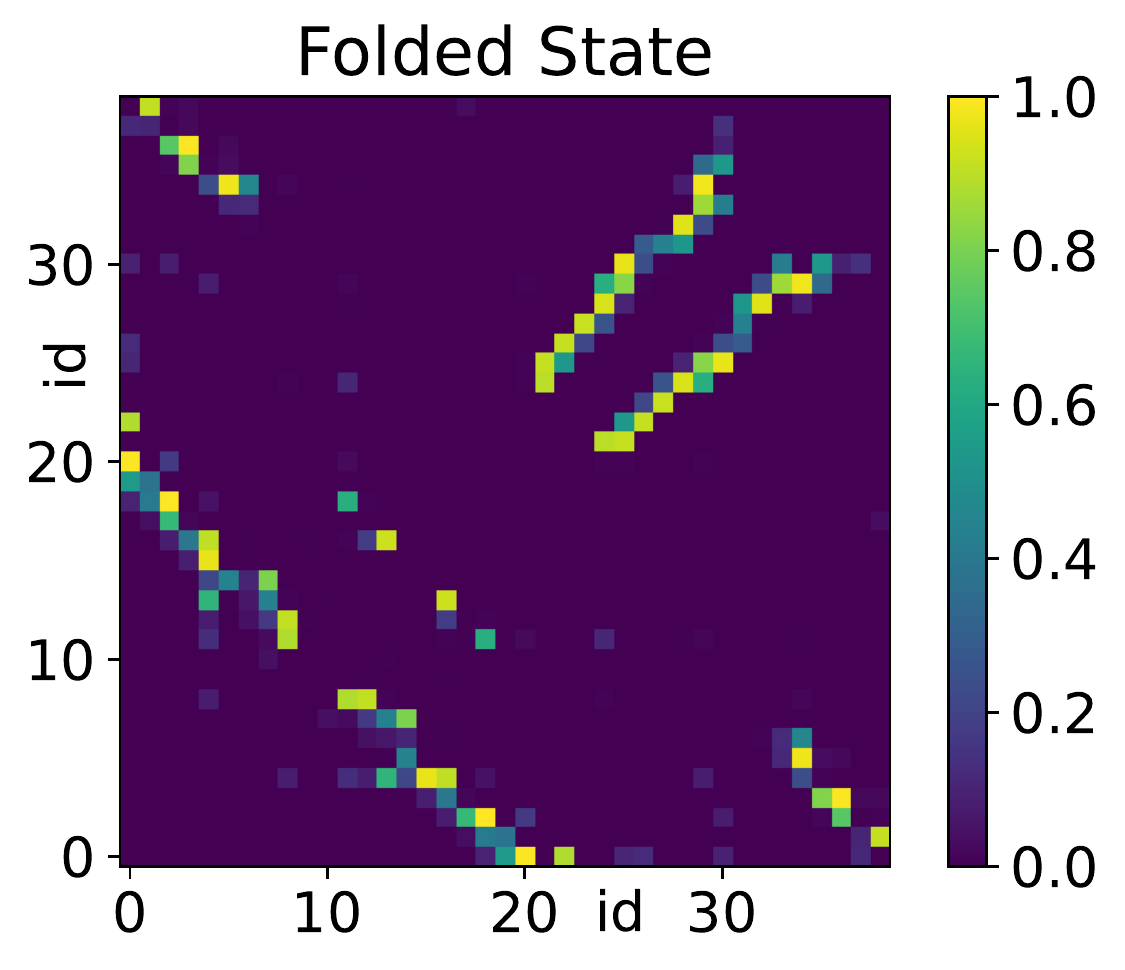}
	\includegraphics[width=0.46\textwidth]{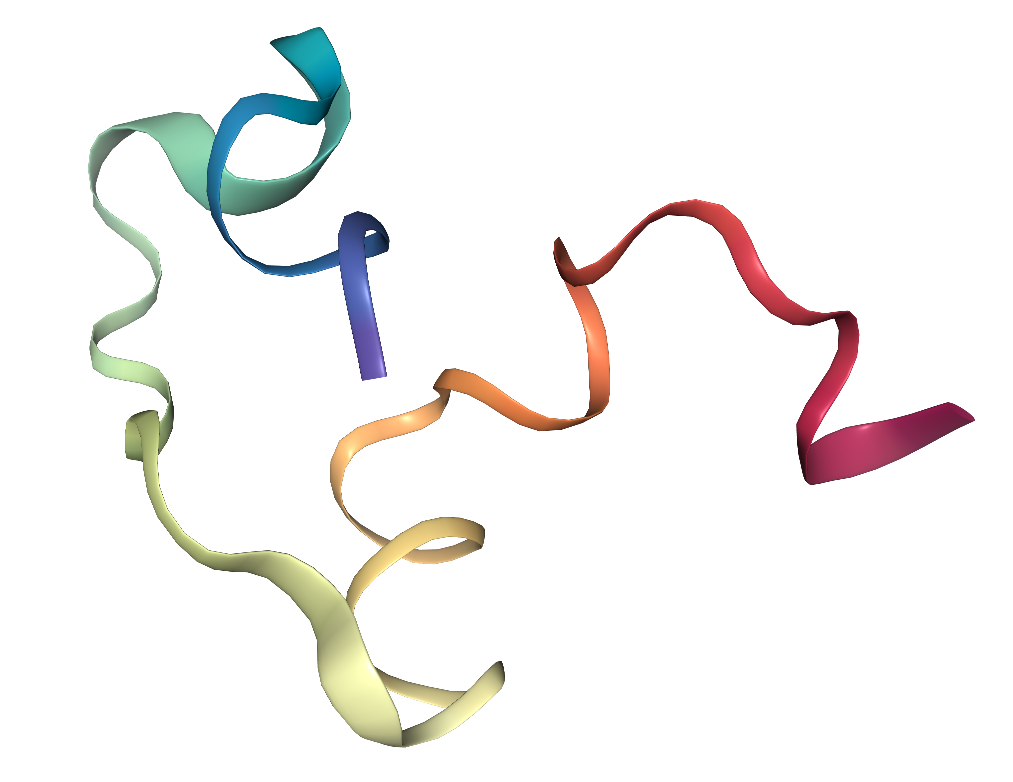}
	\includegraphics[width=0.46\textwidth]{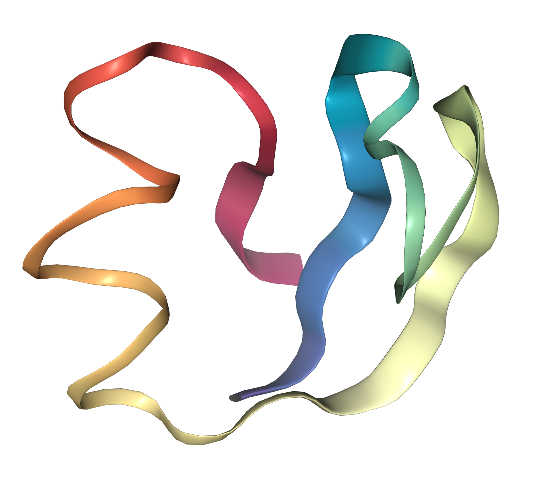}
\caption{Top Row: Representations of the two PCCA states obtained for $\sigma = 15$, $p = 300$, $t = 2\,\mu s$. For each PCCA state, we show the fraction of simulation time during which each residue-residue pair forms a contact, see also~\cite{Nueske2021}. Bottoms Row: Representative protein structure for both PCCA states.}
\end{figure}

\section{Conclusions}

We have introduced a new approach to extract the dominant eigenvalues of dynamical operators associated with stochastic processes, corresponding to the slow timescale dynamics in molecular kinetics. Building on kernel representations of the Koopman operator and generator, our main novelty is a dual low-rank representation of the kernel matrices using random Fourier features. We have derived the corresponding reduced eigenvalue problem, and provided an interpretation of the method in terms of a random dictionary of plane waves. Using four examples of increasing complexity, we have shown that hyper-parameters of the method can be very effectively tuned using the VAMP score metric. Moreover, we have shown that only a few hundred Fourier features were sufficient to obtain accurate and robust results for all systems considered. This finding allowed us to rapidly scan kernel hyper-parameters for data sizes in the ten thousands, on coordinate spaces of dimension up to several hundred. The computational savings  compared to the full eigenvalue problem, measured in floating point operations, were on the order of three to four orders of magnitude.

The main promise of our method is the ability to analyze large data sets using a very generic, and in fact randomly generated basis set. The success on the method depends on the number of required Fourier features. Future research will need to test if this number indeed remains small even for larger systems, which would make the method scalable beyond the benchmarking examples studied here. It will also be interesting to see if kernel functions with more specific properties can be used, for instance, kernels incorporating more detailed symmetries than only translational invariance, or kernels acting on mixed types of domain (periodic and non-periodic). Finally, combining our approach with re-weigthing techniques would allow application of the method to enhanced sampling simulation data.

%\item reviewers: Andy Ferguson, Gabriele Santin, Jerome Henin, Chrstine Peter-Tittelbach

\begin{acknowledgments}
The authors thank D.E. Shaw Research for providing the molecular dynamics simulation data of NTL9.
\end{acknowledgments}

\section*{Data Availability Statement}

The methods presented in this paper are available as part of the KoopmanLib library~\footnote{\url{https://github.com/fnueske/KoopmanLib}}. The complete data and code to reproduce the examples shown above are available from a public repository~\footnote{\url{https://doi.org/10.5281/zenodo.8036525}}, with the exception of the raw molecular dynamics simulation data, which can be obtained directly from the authors or, in the case of NTL9, from the copyright owners.

\bibliography{DMP_Lib}% Produces the bibliography via BibTeX.

\appendix
\newpage
\section{Derivation of Kernel Eigenvalue Problems}
\label{app:derivation_kernel_ev}

Our derivation of the kernel eigenvalue problems in Section~\ref{subsec:rkhs_dyn_op} follows~\cite{Klus2020a,Klus:2020aa}. We compute the matrix elements of the operators $\hat{\mathcal{G}},\,\hat{\mathcal{A}}^t,\,\hat{\mathcal{A}}^L$ on the finite-dimensional space $\bH_m$, with respect to its canonical basis $\{\Phi(x_l)\}_{l=1}^m$:
\begin{align*}
\innerprod{\Phi(x_r)}{\hat{\mathcal{G}}\Phi(x_s)}_\bH &= \frac{1}{m}\sum_{l=1}^m \Phi(x_s)(x_l) \innerprod{\Phi(x_r)}{\Phi(x_l)}_\bH \\
&= \frac{1}{m}\sum_{l=1}^m k(x_s, x_l) k(x_l, x_r) = \frac{1}{m}\left[\K_X \K_X\right](r, s), \\
\innerprod{\Phi(x_r)}{\hat{\mathcal{A}^t}\Phi(x_s)}_\bH &= \frac{1}{m}\sum_{l=1}^m \Phi(x_s)(x_{l+1}) \innerprod{\Phi(x_r)}{\Phi(x_l)}_\bH \\
&= \frac{1}{m}\sum_{l=1}^m k(x_s, x_{l+1}) k(x_l, x_r) = \frac{1}{m}\left[\K_X \K_Y\right](r, s), \\
\innerprod{\Phi(x_r)}{\hat{\mathcal{A}^L}\Phi(x_s)}_\bH
&= \frac{1}{m}\sum_{l=1}^m (\mathcal{L}\Phi(x_s))(x_{l}) \innerprod{\Phi(x_r)}{\Phi(x_l)}_\bH \\
&= \frac{1}{m}\sum_{l=1}^m \left[\nabla F(x_l) \cdot \nabla \Phi(x_s)(x_l) + \frac{1}{2} a(x_l) : \nabla^2 \Phi(x_s)(x_l)\right] k(x_l, x_r) \\
&= \frac{1}{m}\sum_{l=1}^m \left[\nabla F(x_l) \cdot \innerprod{\nabla_1 k(x_l, \cdot)}{\Phi(x_s)}_\bH + \frac{1}{2} a(x_l) : \innerprod{\nabla^2_1 k(x_l, \cdot)}{\Phi(x_s)}_\bH\right]  \\
&\quad  k(x_l, x_r) \\
&= \frac{1}{m}\sum_{l=1}^m \left[\nabla F(x_l) \cdot \nabla_1 k(x_l, x_s) + \frac{1}{2} a(x_l) : \nabla^2_1 k(x_l, x_s)\right]  k(x_l, x_r) \\
&= \frac{1}{m}\left[\K_X \K_X^L\right](r, s).
\end{align*}
Here, we used the subscript $\nabla_1$ for the nabla-operator to indicate differentiation with respect to the first variable. To get from the second to the third line, we used the derivative reproducing property
\[ D^\alpha f(x) = \innerprod{D^\alpha_1 k(x, \cdot)}{f}_\bH, \]
which holds for any multi-index $\alpha$ and all RKHS functions $f$ as soon as the kernel is at least $2|\alpha|$-times continuously differentiable in both arguments, see~\cite{WENDLAND04}.

We see that in all cases, the full-rank matrix $\K_X$ can be canceled out, which leaves us with~\eqref{eq:ev_problem_kernel}. In the reversible case, we replace the operator $\hat{\mathcal{A}}^L$ by a symmetric bi-linear form
\[ \hat{\mathcal{A}}^\mathrm{rev}(f, g) = -\frac{1}{2m}\sum_{l=1}^m \nabla^\top f(x_l) a(x_l)\nabla g(x_l). \]
Computing its matrix elements on $\bH_m$, we find
\begin{align*}
\hat{\mathcal{A}}^\mathrm{rev}(\Phi(x_r), \Phi(x_s)) &= -\frac{1}{2m}\sum_{l=1}^m \nabla^\top \Phi(x_r)(x_l) a(x_l)\nabla \Phi(x_s)(x_l) \\
&= -\frac{1}{2m}\sum_{l=1}^m \nabla_1^\top k(x_l, x_r) a(x_l) \nabla_1 k(x_l, x_s) = \K_X^\mathrm{rev}(r, s).
\end{align*}
In this case, we cannot cancel a factor $\K_X$, which leads to~\eqref{eq:kernel_ev_rev}.

\end{document}